\documentclass[reprint,twocolumn]{revtex4-1}

\usepackage[english]{babel}
\usepackage{makeidx}
\usepackage{bm, amsmath, amssymb}
\usepackage{graphicx,subfigure}
\usepackage{amsfonts}
\usepackage{amsmath}
\usepackage{amssymb}
\usepackage{graphicx}
\usepackage[cp1251]{inputenc}
\usepackage[T2A]{fontenc}
\usepackage{color}
\DeclareMathOperator{\csgn}{csgn}
\DeclareMathOperator{\sgn}{sgn}
\DeclareMathOperator{\Tr}{Tr}

\begin{document}
\title{Analytic density of states of \textcolor{black}{tight-binding model} for two-dimensional  Chern insulator}

\author{Vera Uzunova}
\affiliation{Institute of Theoretical Physics, Faculty of Physics, University of Warsaw, ul. Pasteura 5, 02-093 Warszawa, Poland}
\affiliation{Institute of Physics of the National Academy of Sciences of Ukraine, 46 Nauky Ave., 03039 Kyiv, Ukraine}

\author{Krzysztof Byczuk}
\affiliation{Institute of Theoretical Physics, Faculty of Physics, University of Warsaw, ul. Pasteura 5, 02-093 Warszawa, Poland}

\date{\today}

\begin{abstract}

We present analytic expressions for the density of states and its consistent derivation for the two-dimensional Qi-Wu-Zhang (QWZ) Hamiltonian, a generic model for the  Chern topological insulators of class A. This density of states is expressed in terms of  elliptical integrals. We discuss and plot special cases of the dispersion relations and the corresponding densities of states. Spectral moments are also presented. The exact formulae ought to be useful in determining physical properties of the non-interacting Chern insulators and within the dynamical mean-field theory  for interacting fermions with the QWZ Hamiltonian in the non-interacting limit.  

\end{abstract}

\maketitle

\section{Introduction}

A topological insulator (TI) is a common name for the novel class of systems with non-trivial topological properties \cite{Kane,Zhang}. Historically, the first example of TI was a two-dimensional electron gas in a strong magnetic field where the integer quantum Hall effect was observed \cite{Laughlin}. After predicting and later discovering of other examples of TIs \cite{nature-mat} the subject becomes a main stream of condensed matter physics \cite{cond-mat}, of cold atoms in optical lattices \cite{cold}, of photonics \cite{photonics}, and even of electric engineering \cite{PT}. 

One of a possible path to investigate TIs is to study tight-binding models defined on particular lattices. Among various interesting examples are either the Su-Schrieffer-Heeger model \cite{SSH} and the Rice-Mele model \cite{Rice} in one-dimension or the Haldane model \cite{Haldane} and the Qi-Wu-Zhang (QWZ) model in two dimensions \cite{QWZ}. The latter one is defined on a square lattice and the corresponding two dimensional Brillouin zone whereas the former one is formulated on a hexagonal lattice. 

In particular, the QWZ model is a well-known\textcolor{black}{two-band} system in studying physics of fermions such as bulk and edge  properties, different topological states,  thermodynamics, transport,  and many others \cite{Short}. This model is also used as a non-interacting part of the many-body Hamiltonian where the two-particle interaction is included. Its further generalization to arbitrary dimensions and even to the limit of infinite dimension proved that topological insulators are possible in interacting systems as well \cite{Potthoff}. 

In spite of such a broad interest in the QWZ model its density of states (DOS) is  not yet determined analytically. Although the DOS by itself is not sufficient to provide topological classification of a system, it is a basic and very important quantity necessary to investigate thermodynamics, thermodynamic phases, response of the system on different probes, and many other quantities.\textcolor{black}{However, its derivative, cf. Streda formula \cite{Streda}, can serve as a topological indicator.} 

To fill in this gap, in this article we derive analytical formulae of the DOS in terms of\textcolor{black}{special functions known as complete elliptic integrals, which mathematical properties are defined and tabulated \cite{abramowitz}}. We discuss details of the derivation and basic properties of the DOS when the relevant model parameter,\textcolor{black}{a mass term controlling the topology}, is varied.\textcolor{black}{Our results are exact, analytic and free of any numerical inaccuracy. They allow for  a complete and comprehensive understanding of the physics (thermodynamics, dc transport, cf. STM current, or spectral properties) of the QWZ model, which up to now was mostly investigated from  the point of view of topological properties. Since the QWZ model plays so important role in the field, the necessity of its holistic understanding cannot be overestimated. } 

The DOS, denoted here by $\rho(\Omega)$, counts the number of states in a vicinity of a particular value of energy $\Omega$, i.e. $dN=\rho(\Omega) d\Omega$. It can be obtained from the\textcolor{black}{single-particle} Green's function\textcolor{black}{(resolvent)}.  Analytic derivation  of the DOS, even for two-dimensional (2D) systems, is typically a challange and thus there are  very few known analytical results. One of the first example was obtained in 1953 by Hobson and Nierenberg \cite{Hobson}. It is an analytical expression for the DOS of graphene with the nearest-neighbor hopping and represented by the complete elliptical integrals. The consequent derivation of this result is presented in \cite{Ovchynnikov}. The DOS for some others 2D latices are also obtained analytically \cite{Kogan}. In particular, it can be done for square, triangular, honeycomb, Kagome, and Lieb lattices. In this paper the DOS is  analytically derived for the QWZ\textcolor{black}{two-band} Hamiltonian modeling a Chern insulator on the square lattice. 
\textcolor{black}{In contrast to earlier examples, our analytic DOS depends explicitly on a mass parameter, which allows to change the topology of the  sytem. In fact, we derived a whole family of DOS in exact, analytic forms for this two-band QWZ system in two dimensions. 
For  three dimensional tight-binding models the densities of states  in analytic forms were determined for a simple cubic lattice, a body center lattice, and for a face centered lattice \cite{Jelitto68,Sakaji02,Hijjawi04}. }

\textcolor{black}{Knowledge of the DOS in analytic form is invaluable in further investigation of physical properties of tight-binding model. Such analytic form allows for a precise determination of van Hove singularity positions and their types inside the DOS. And, as a result, it helps to achieve high  accuracy when integrals 
involving the DOS are performed numerically. }

\textcolor{black}{In certain numerical applications analytic forms of the DOS improves numerical efficiency of programs. For example, in the dynamical mean-field theory (DMFT) the  semi-elliptic DOS, given analytically, allows to determine the Hilbert transform exactly and this simplifies the self-consistency equations \cite{Georges96}. Also, the study of fermions in the infinite dimensional limit within the DMFT used the analytic Gaussian form of the DOS on hyper-cubic lattice \cite{Georges96} to investigate metal-insulator and antiferromagnetic transitions, and the analytic expression of the DOS for face-centered lattice in the same dimension limit \cite{Ulmke98} to investigate itinerant ferromagnetism. Following the latter, the hand tailored analytic DOS with a free parameter controlling its asymmetry was used within DMFT to study detailed conditions for occurring the itinerant ferromagnetism inside a single band Hubbard model \cite{Wahle98}. A recent study of an extended Falicov-Kimball model also used different DOS provided in simple analytic forms  \cite{Kapcia21}. Since analytic forms of the DOS are so important, some DOS for selected lattices were included explicitly in terms of elliptic integrals inside the  Python library devoted to Green's functions in tight-binding models \cite{python}. 
}

Our presentation is organized as follows: In section II we define the QWZ model and discuss the dispersion relations, in Section III we introduce the DOS and present its analytic derivation for the QWZ Hamiltonian, Section IV is devoted to the discussion of the  DOS in different parameter regimes, in Section V we show  some additional features in the DOS, the spectral moments are discussed in Section VI, and we close our presentation with  Section VII, where we offer our conclusions and  outlooks. In  Appendix A we provide mathematical definitions of the elliptic integrals and in the Appendix B   we give selected details on calculating the spectral moments. 

\section{QWZ model Hamiltonian in two dimensions}

A generic form of the two-band Hamiltonian for a 2D noninteracting system in the momentum space can be written as
\begin{equation}\label{H}
\hat{H}= \sum_{\bf k} \hat{H}_{\bf k} =\sum_{\bf k} \mathbf{h}({\bf k})\cdot \hat{\boldsymbol{\sigma}},
\end{equation}
where $\mathbf{k}=(k_x,k_y)$ ($-\pi/a_L <$ $k_x,$ $k_y$ $\leq \pi/a_L$)  is a 2D wave vector in the first Brillouin zone corresponding to the 2D square lattice with the lattice constant $a_L$, ${\bf h}({\bf k})$ is a vector with three components being given functions of $\bf k$, and 
$\hat{\boldsymbol{\sigma}}$ is the vector with components represented by  the three Pauli matrices $\hat{\sigma}_x$, $\hat{\sigma}_y$, and $\hat{\sigma}_z$. 

The Hamiltonian (\ref{H}) describes a two level system, corresponding to either the two orbitals or the spin $1/2$ degrees of freedom. The vector 
${\bf h}({\bf k})$ is interpreted as a Zeeman like magnetic field with some (perhaps non-trivial) dependence on the wave vector $\bf k$. 
This model breaks the time reversal symmetry and belongs to the class A in the ten fold way classification scheme \cite{ten-way}. 
The Hamiltonian (\ref{H}) can be easily diagonalized, giving a two band energy spectrum $\epsilon_{\pm}(\mathbf{k})=\pm h({\bf k}) $, where 
$h({\bf k})=|{\bf h}({\bf k})|$  is the length of the vector $\mathbf{h}(\mathbf{k})$. 

In the following we consider a particular parametrization  where the  length of ${\bf h}({\bf k})$  is given by $h({\bf k})^2=m^2+2t^2+2t^2\cos (k_xa_L) \cos (k_y a_L)+2 m t [\cos (k_xa_L)+\cos (k_ya_L)]$. It corresponds to the following representation  of the vector $\mathbf{h}({\bf k})$:
\begin{align}
h_x({\bf k})=t\sin (k_xa_L),\notag\\
h_y({\bf k})=t\sin( k_ya_L), \notag \\
h_z({\bf k})=m+t\cos (k_xa_L)+t\cos (k_ya_L),
\end{align}
where $t$ is the hopping amplitude. In the momentum space this vector $\mathbf{h}(\mathbf{k})$ has a Skyrmion configuration for $0<|m|/t<2$ \cite{Short}. In other words, the system is a topological insulator with the finite Chern number $\pm 1$ and conducting surface states at the half-filling. Hamiltonian (\ref{H}) can be interpreted as a tight-binding model of a magnetic semiconductor with the Rashba-type spin-orbit interaction and a uniform magnetization along the $z$-axis \cite{Short}.  
In the following we take $t=1$, which sets the energy unit. We also use $a_L=1$ for the length unit. 

\begin{figure}[h!]
\centering
\includegraphics[width=0.45\textwidth]{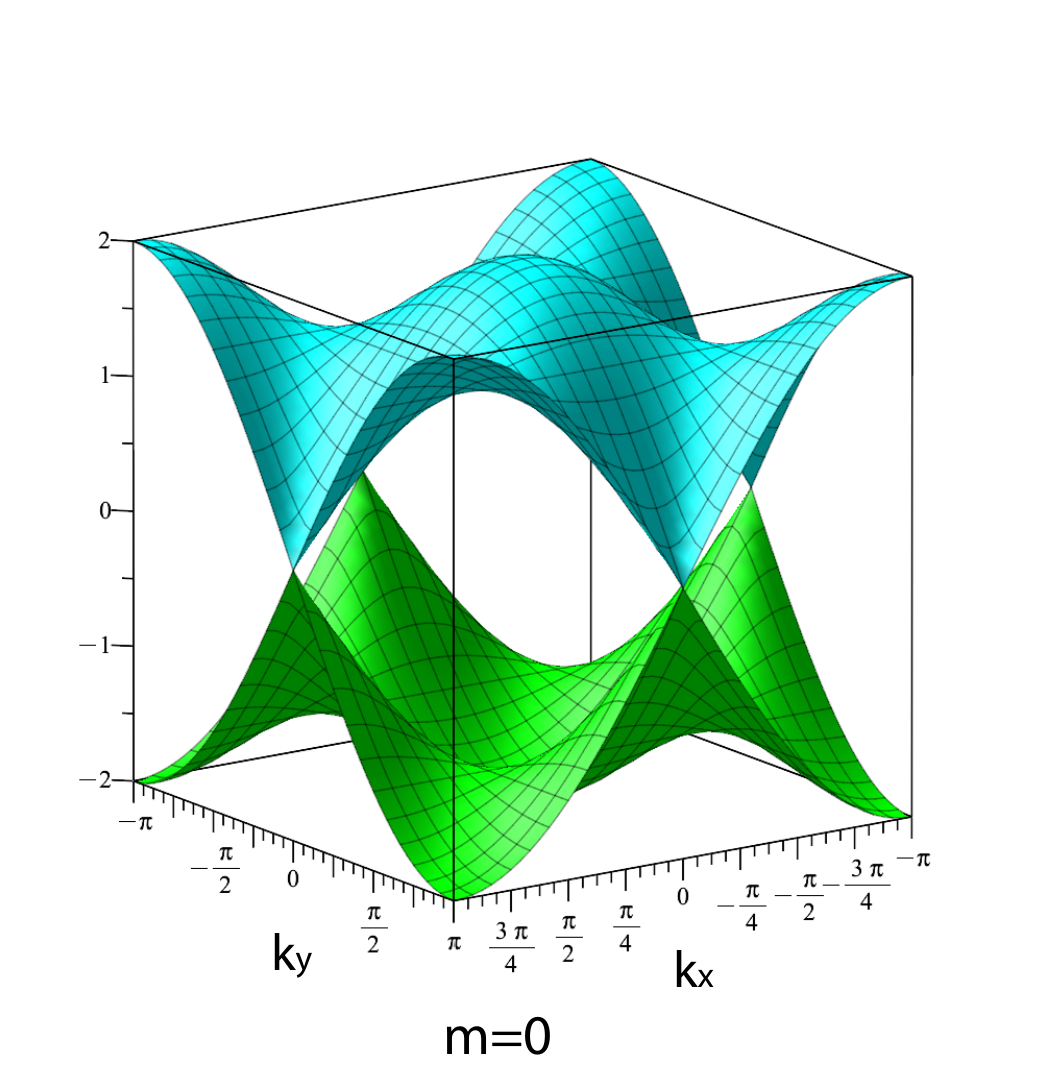}
\caption{ Dispersion relations $\epsilon_{\pm}(\mathbf{k})$ of the QWZ model at $m=0$. The gap is closed at $X=(\pm\pi,0)$ and $Y=(0,\pm\pi)$ points.}
\label{disp0}
\end{figure}

\begin{figure}[h!]
\centering
{\includegraphics[width=0.45\textwidth]{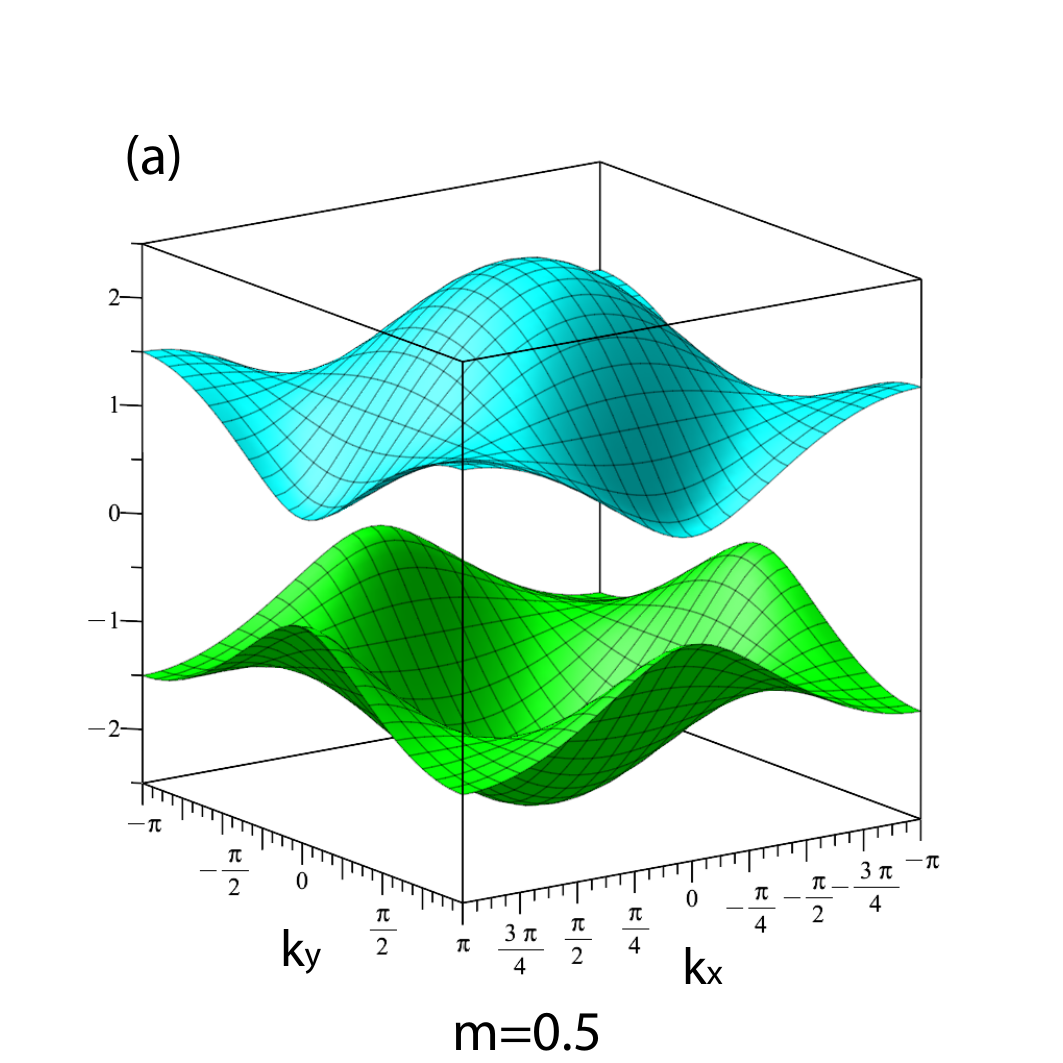}}
{\includegraphics[width=0.45\textwidth]{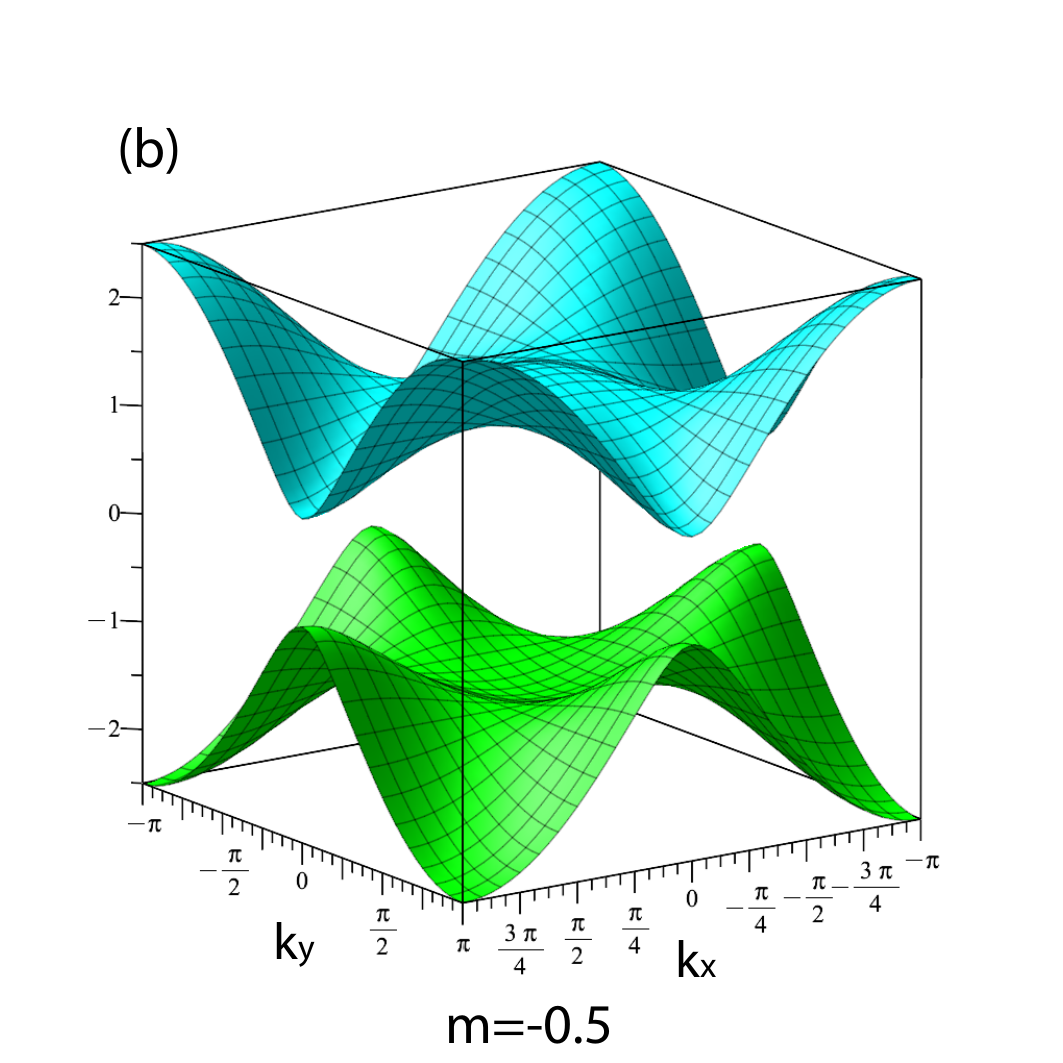}}
\caption{Dispersion relations $\epsilon_{\pm}(\mathbf{k})$ of the QWZ model at: (a) $m=0.5$, and (b) $m=-0.5$. }
\label{disp05} 
\end{figure}

\begin{figure}[ht!]
\centering
{\includegraphics[width=0.45\textwidth]{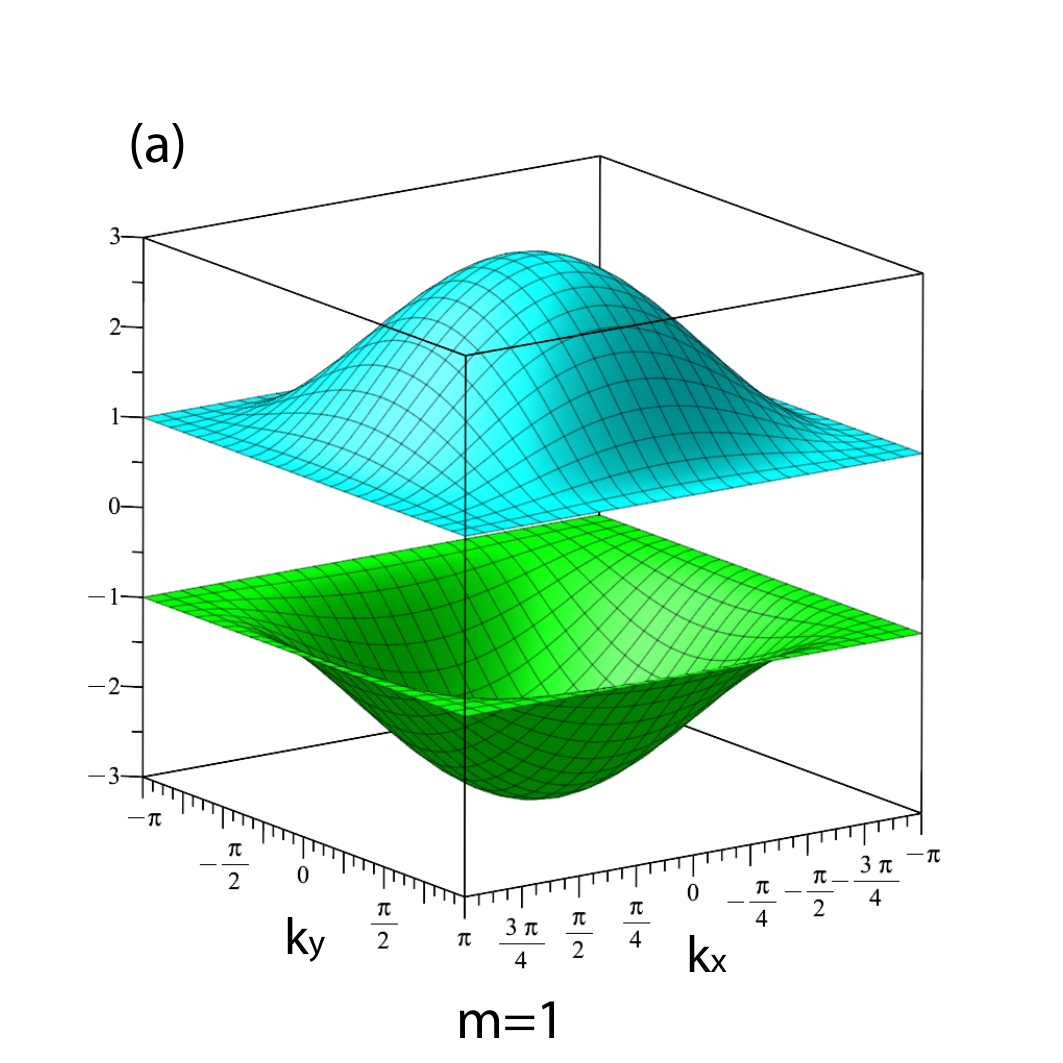}}
{\includegraphics[width=0.45\textwidth]{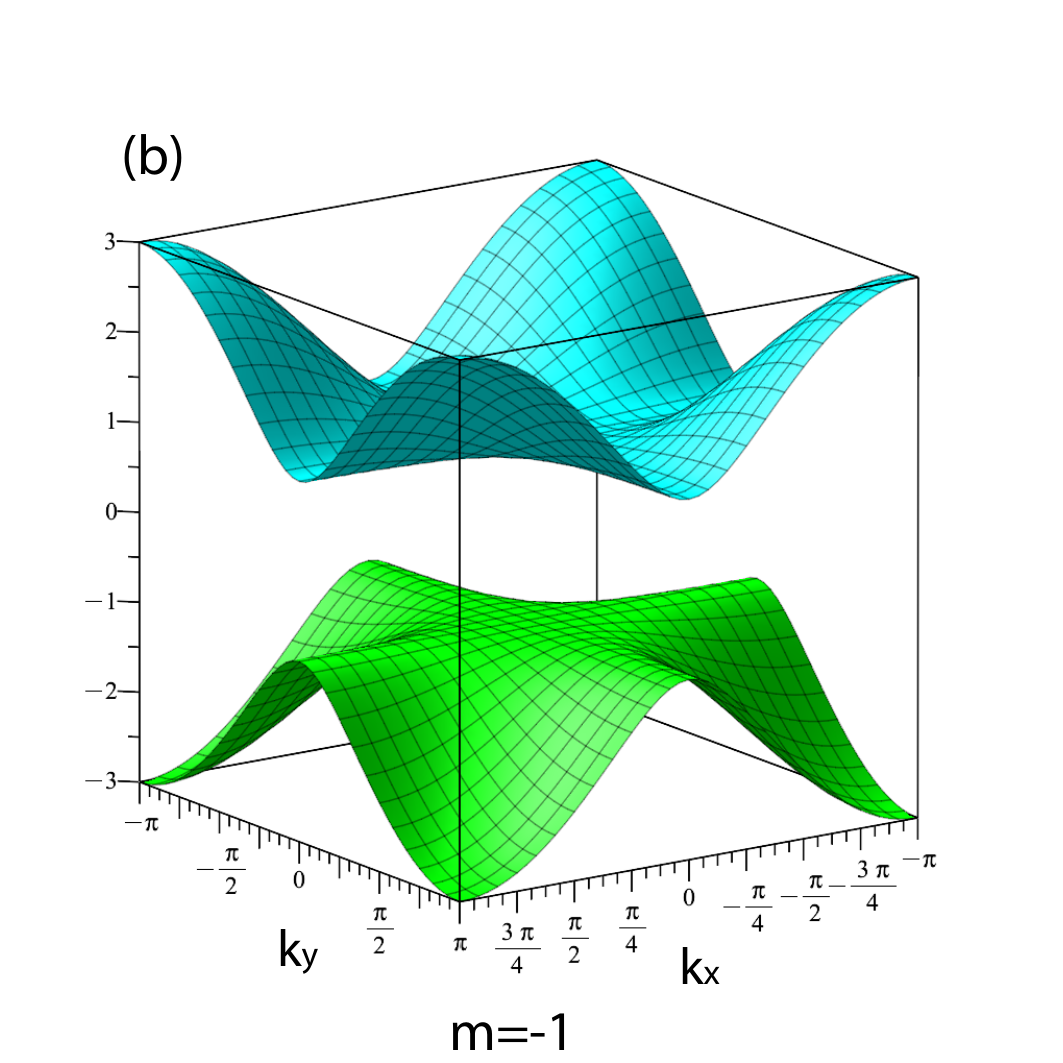}}
\caption{\label{disp1} Dispersion relations $\epsilon_{\pm}(\mathbf{k})$ of the QWZ model at: (a) $m=1$, and (b) $m=-1$. }
\end{figure}

\begin{figure}[ht!]
\centering
{\includegraphics[width=0.45\textwidth]{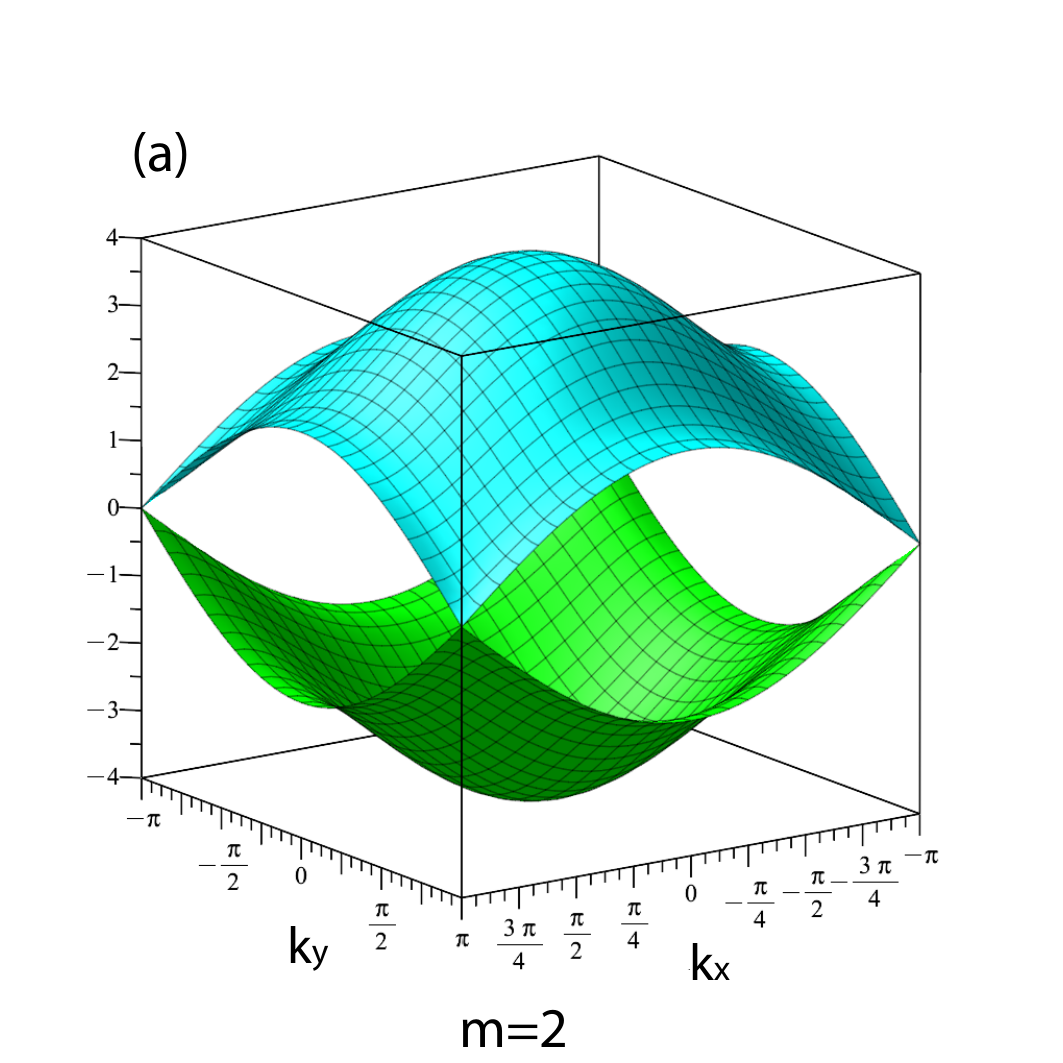}}
{\includegraphics[width=0.45\textwidth]{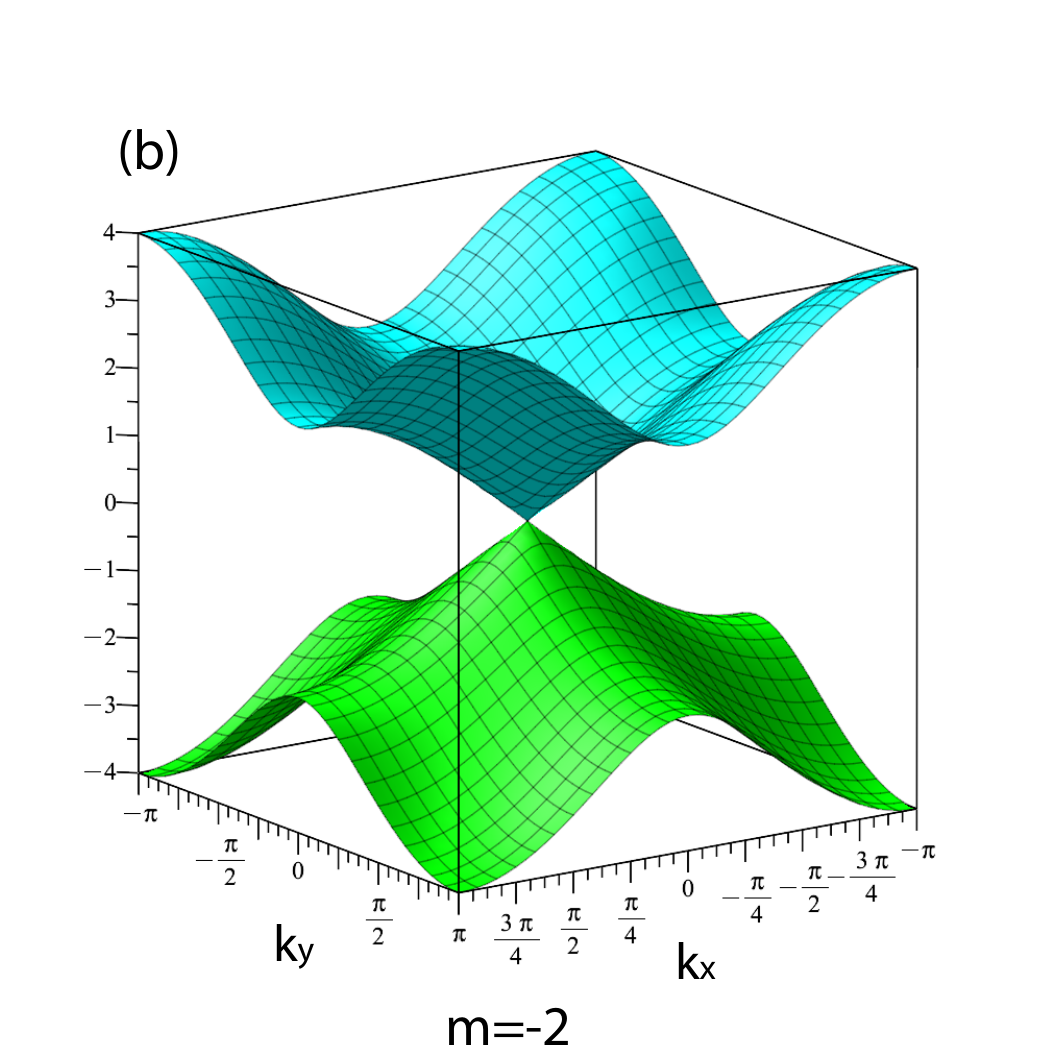}}
\caption{ Dispersion relations $\epsilon_{\pm}(\mathbf{k})$ of the QWZ model at: (a) $m=2$, and (b) $m=-2$. For $m=2$ the gap is closed at $M=(\pm\pi,\pm\pi) $ and $(\pm\pi,\mp\pi)$ points and for $m=-2$ it is closed at $\Gamma=(0,0)$ point. }
\label{disp2}
\end{figure}

\begin{figure}[ht!]
\centering
\includegraphics[width=0.45\textwidth]{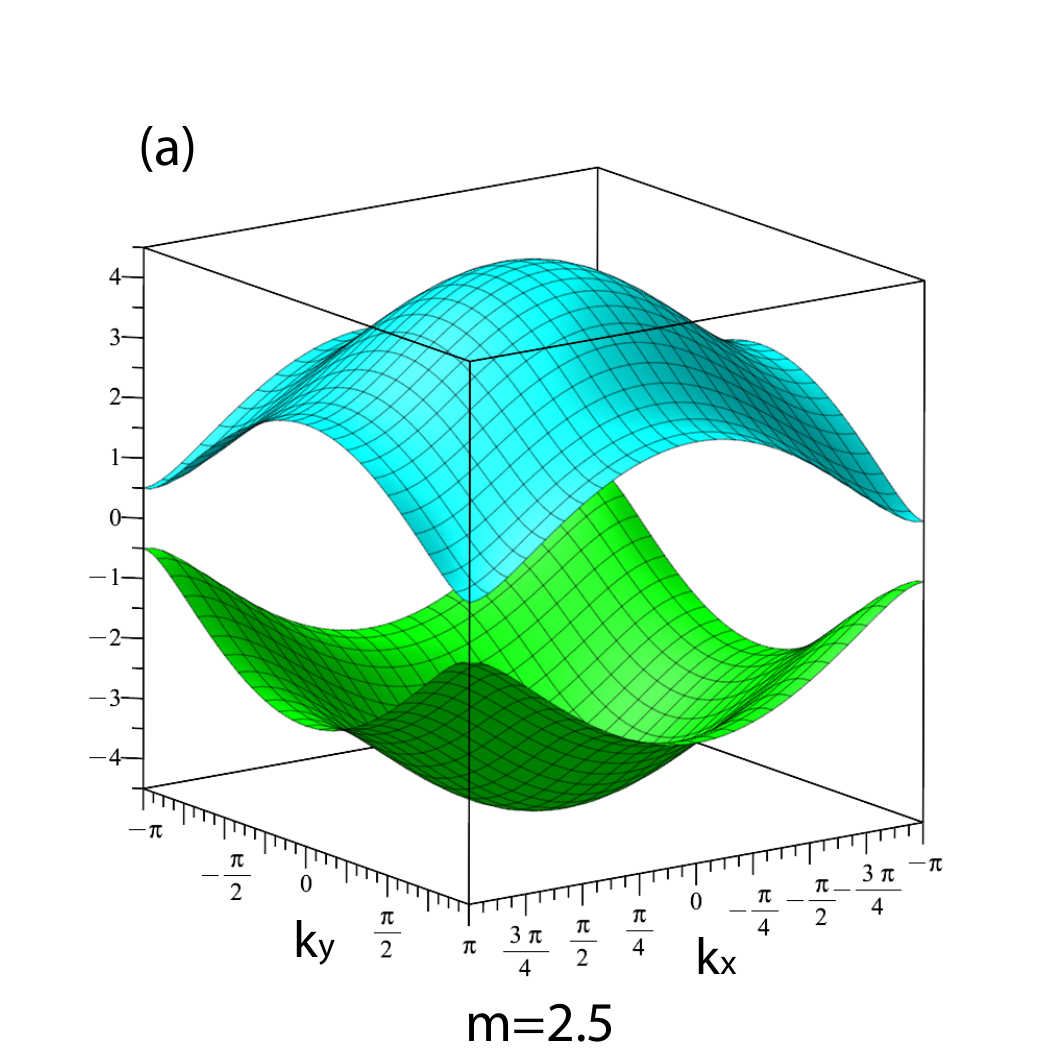}
\includegraphics[width=0.45\textwidth]{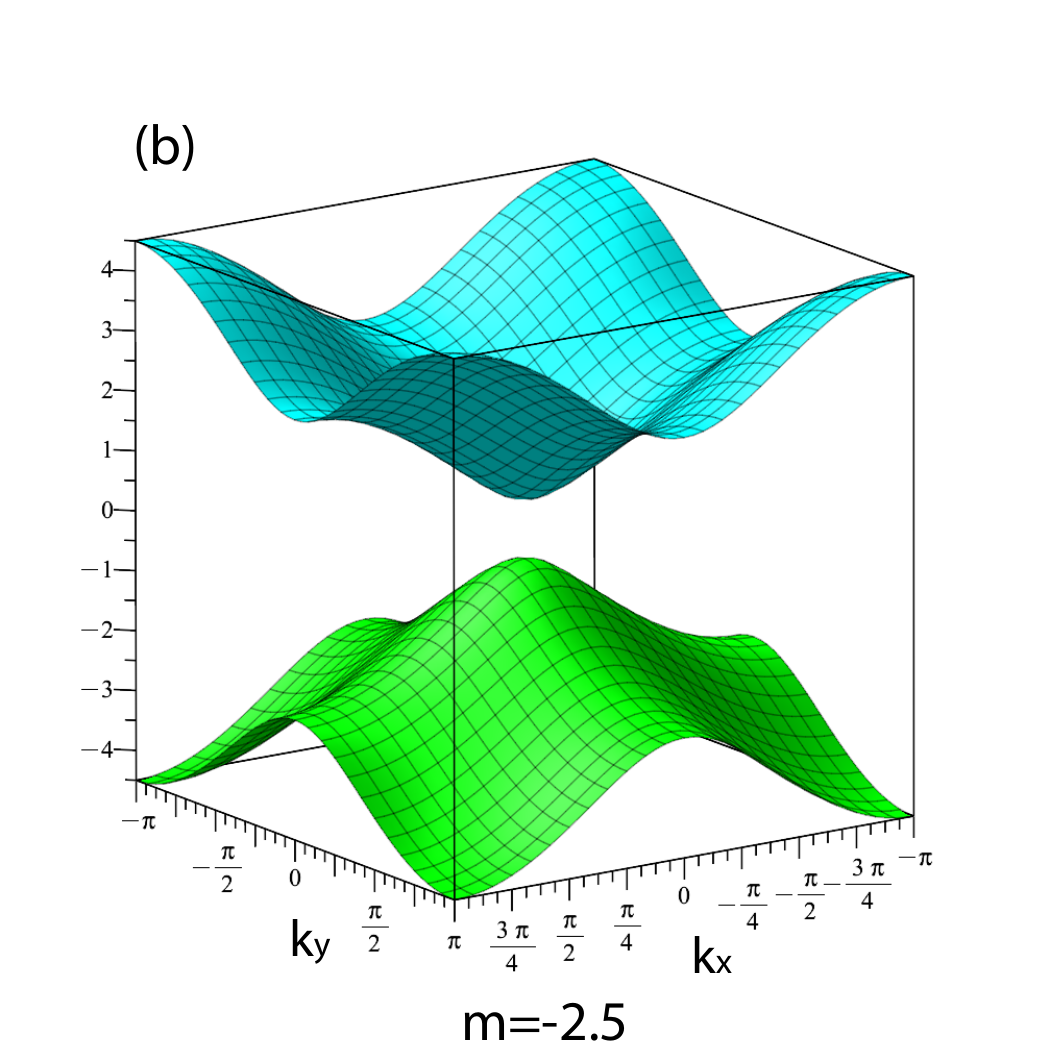}
\caption{ Dispersion relations $\epsilon_{\pm}(\mathbf{k})$ of the QWZ model at: (a) $m=2.5$, and (b) $m=-2.5$. }
\label{disp25}
\end{figure}

For $m=0$, $0.5$, $1$, $2$, and $2.5$  the corresponding eigenvalues (energy bands) of the Hamiltonian~(\ref{H}) are plotted in Figs.~\ref{disp0}-\ref{disp25}, respectively. We see that at $m=0$ (Fig.~\ref{disp0}) and $m\pm2$ (Fig.~\ref{disp2}) the band gap is closed at $\Gamma$, $X$ ($Y$) and $M$ special points in the square Brillouin zone, respectively, and characteristic Dirac cone(s) is (are) formed. For $m\neq 0$, $\pm 2$ the bands are separated by the gap as seen in Figs.~\ref{disp05}, \ref{disp1}, and \ref{disp25}. 

\widetext

\endwidetext

\section{Analytic derivation of DOS} 

\subsection{DOS and its symmetry}

The band resolved DOS per lattice site for the lower $\epsilon_{-}(\mathbf{k})=- h({\bf k}) $ (upper $\epsilon_{+}(\mathbf{k})=+ h({\bf k}) $) band is defined as usual 
\begin{equation}
\rho_{\pm}(\Omega) = \frac{1}{N_L} \sum_{\bf k} \delta (\Omega -\epsilon_{\pm}(\mathbf{k})), 
\end{equation}
where $\delta (x)$ is the Dirac distribution function and $N_L$ is the number of lattice sites. Due to the symmetry of the Dirac function one can easily show that 
\begin{equation}
\rho_+(\Omega)=\rho_-(-\Omega). 
\end{equation}
The total DOS per lattice site is defined as a sum of them, i.e., 
\begin{equation}
\rho(\Omega) = \rho_+(\Omega) + \rho_-(\Omega) .
\end{equation}
Since the lower and upper bands do not overlap \cite{footnote} the band resolved DOS can be directly extracted form the total DOS, i.e., 
\begin{equation}
\rho_{\pm}(\Omega)=  \theta(\pm \Omega) \rho(\Omega) , 
\end{equation}
where $\theta(x)$ is a step function. Therefore, in the following we focus on deriving a formula for the total DOS. 

\subsection{Retarded Green's function and total DOS}

Equivalently, the total DOS per lattice site is determined  as a trace in band indices ($\pm$) of the imaginary part of the retarded Green's function 
\begin{equation}\label{rho0}
\rho(\Omega) =-\frac{1}{\pi }\Im [\Tr \mathbf{G}(\Omega) ].
\end{equation}
The diagonal matrix elements of the Green's function $\mathbf{G}(\Omega) $ are
\begin{equation}
G_{\pm}(\Omega)=\frac{1}{N_L} \sum_{\mathbf{k}}\frac{1}{\Omega-\epsilon_{\pm}(\mathbf{k}) +\imath 0^+},
\end{equation} 
and for a while the small imaginary part $\imath 0^+$ will not be written explicitly.  
Summation over $\textbf{k}$ can be replaced by the continuous integral in the first Brillouin zone $\sum_{\mathbf{k}}\rightarrow \frac{L^2}{(2\pi)^2} \int_{BZ} dk_x dk_y$, where $L$ is the length of the system and $L^2=N_La_L^2$, with $a_L=1$.
Then the trace of the Green's function has the form
\begin{equation}
\Tr \mathbf{G}(\Omega) =\frac{\Omega}{2\pi^2}\int_{-\pi}^{\pi}\int_{-\pi}^{\pi}\frac{dk_x dk_y}{\Omega^2-h({\bf k}) ^2}.
\end{equation}

Integrations with respect to $k_x$ and $k_y$ are symmetric. We first preform the integration of the function $(\Omega^2-h(k_x,k_y)^2)^{-1}$ with respect to $k_x$. It is convenient to denote
\begin{align}\label{iab}
a=\Omega^2-(m^2+2+2m\cos k_y),\notag\\
b=-2(m+\cos k_y),
\end{align}
 and then to use the identity (2.558.4) in \cite{Gradshteyn}
\begin{align}\label{Gr0}
\int\frac{d k_x}{a+b\cos k_x}=\frac{2\pi}{\sqrt{a^2-b^2}}\arctan\left[\frac{a-b}{\sqrt{a^2-b^2}}\tan\left(\frac{k_x}{2}\right)\right],
\end{align}
where $a^2-b^2>0$. For the case $a^2-b^2\le 0$ the result of the integration (\ref{Gr0}) can be formally rewritten as 
\begin{equation}\label{Ix}
\int_{-\pi}^{\pi}\frac{dk_x}{\Omega^2-h^2}=\frac{-2\pi\imath}{\sqrt{b^2-a^2}}\csgn\left[\frac{\imath(b-a)}{\sqrt{b^2-a^2}}\right],
\end{equation}
where it is taken into account that $\arctan(\pm\tan\pi/2)=\pm\pi/2$. The complex signum function, abbreviated as  $\csgn$, is equal to the sign (function $\sgn$) of the real part of the argument, and sign of the imaginary part if the real part is zero.  
We analytically continue this result by taking $\Omega\rightarrow \Omega+\imath 0^{+}$, where we add infinitesimally small imaginary part as it is required in the retarded Green's function. 
Then $a=(\Omega+\imath 0^{+})^2+...$ gets small imaginary part $2 \imath \Omega0^{+}$ defining the value of $\csgn$. Taking the limit of infinitisimal $0^{+}$ we arrive at  the following expression for the trace of the Green's function 
\begin{equation}\label{Iy}
\Tr \mathbf{G}(\Omega) = \frac{2\Omega}{\pi}\int_0^\pi d k_y
 \begin{cases}
   \frac{\sgn{a}}{\sqrt{a^2-b^2}} & \text{for} \;\; a^2>b^2,\\
   \imath\frac{ \sgn{\Omega}}{\sqrt{b^2-a^2}} & \text{for} \;\; a^2\leq b^2,
 \end{cases}
\end{equation}
where we used the fact that the integrand is an even function of $k_y$. Here the values $a$ and $b$ are the functions of the variables $k_y$ and $\Omega$ and the parameter $m$ as determined by Eqs.~(\ref{iab}).

The DOS is proportional to the imaginary part of $\Tr G(\Omega)$, therefore it is nonzero only in the region of $\Omega$ where
\begin{equation}\label{ab}
a^2-b^2\leq 0.
\end{equation}
Outside the region (\ref{ab}) $\Tr G$ is real and the DOS is equal to zero, i.e., $\rho(\Omega) =0$. 

The next step in evaluation of Eq.~(\ref{Iy}) is to perform integration over $k_y$. It is convenient to replace $y=\cos k_y$ and $d k_y=-dy (1-y^2)^{-1/2}$. The boundaries of the integration are $-1\leq y \leq 1$, where changing the order of the integration boundaries results an additional minus sign. The integration over $y$ is performed differently depending on the value of the parameter $m$. Therefore,  in what follows we are considering three cases with $|m|=1$, $|m|>1$, and $|m|<1$, separately. 

\subsubsection{Case $|m|=1$}

Calculations are simpler in the special case of $|m|=1$. The function in the denominator of the integral~(\ref{Iy}) can be written explicitly as $a^2-b^2=(\Omega^2-1)(\Omega^2-5-4my)$. It is linear in $y$ and changes the sign only once at the point $y=y_0 \sgn{m}$, where $y_0=(\Omega^2-5)/4$. Solving the condition (\ref{ab}) with respect to $\Omega$ and using the fact that $|y|\le 1$ we obtain that DOS is nonzero only for $1\le |\Omega|\le 3$.   

The condition (\ref{ab}) considered with respect to $y$ determines the boundaries of integration in Eq.~(\ref{Iy}). For $m=1$ it is satisfied for $y_0\le y\le 1$.Then the imaginary part of Eq.~(\ref{Iy}) in terms of variable $y$ gives the DOS in the implicit form 
\begin{align}\label{r01}
\rho(\Omega)=\frac{|\Omega|}{\pi^2\sqrt{\Omega^2-1}}\int_{y_0}^1 \frac{dy}{\sqrt{(1-y^2)(y-y_0)}},
\end{align}
where the definition~(\ref{rho0}) is used. 
The similar expression will appears for $m=-1$: the boundaries of integration are $-1 \le y\le -y_0$ and denominator in the integrand is $\sqrt{(1-y^2)(-y-y_0)}$. The corresponding DOS can be transformed into the result Eq.~(\ref{r01}) by replacing the integration variable $y\rightarrow -y$. So Eq.~(\ref{r01}) is valid for both cases $m=\pm 1$.

The integral over $y$ can be calculated in terms of the complete elliptical integral of the first kind $K(x)$ using the identity (3.131.5) from \cite{Gradshteyn}
\begin{align}
\int_{u_2}^{u_3}\frac{dy}{\sqrt{(u_3-y)(y-u_2)(y-u_1)}}\notag\\
=\frac{2}{\sqrt{u_3-u_1}}K\left[ \sqrt{\frac{(u_3-u_2)}{(u_3-u_1)}}\right],
\end{align}
where $u_3>u_2>u_1$. In Eq.~(\ref{r01}) these parameters are $u_1=-1$, $u_2=y_0$, $u_3=1$ and the result of integration is equal to $\sqrt{2}K\left[\sqrt{(1-y_0)/2}\right]$.
Then the DOS for $|m|=1$ is  
\begin{equation}\label{r1}
\rho_{|m|=1} (\Omega) =  \begin{cases}
   \frac{\sqrt{2}}{\pi^2}\frac{|\Omega|}{\sqrt{\Omega^2-1}}K\left[\sqrt{\frac{9-\Omega^2}{8}}\right],\;\;\;\;\text{if}\;\; 1\leq\Omega^2\leq 9,\\
   0, \;\; \;\;\;\;\;\;\;\;\;\;\;\;\;\;\;\;\;\;\;\;\;\;\;\;\;\;\;\;\;\;\;\;\;\;\;\text{otherwise}.
 \end{cases}
\end{equation}
For convenience, in the Appendix A we provide the definitions of the elliptic integrals.

\subsubsection{Case $|m|>1$}

To calculate the DOS for all other values of the parameter $m$ we need first to analyze the function in denominator of Eq.~(\ref{Iy}). For $|m|\neq 1$ we can write $a^2-b^2=4(m^2-1)(y-y_1)(y-y_2)$. Here the left and right zeros of the function are denoted as $y_1=\min (\tilde{y}_1,\tilde{y}_2)$ and $y_2=\max (\tilde{y}_1,\tilde{y}_2)$, respectively, where
\begin{align}\label{y12}
\tilde{y}_1=\frac{\Omega^2-1-(m+1)^2}{2(m+1)}, \notag \\
\tilde{y}_2=\frac{\Omega^2-1-(m-1)^2}{2(m-1)}.
\end{align}
So we have to set  $y_1=\tilde{y}_1$, $y_2=\tilde{y}_2$ if the the following condition is met 
\begin{align}\label{yy}
\frac{\Omega^2+m^2-2}{m^2-1}>0,
\end{align}
and to set $y_1=\tilde{y}_2$, $y_2=\tilde{y}_1$, otherwise.

Lets consider the case $|m|>1$, then factor $m^2-1$ is positive. Rewriting Eq.~(\ref{Iy}) in terms of variable $y$ and substituting it in Eq.~(\ref{rho0}), we obtain the nonzero DOS in the 
implicit form
\begin{align}
\rho(\Omega)=\frac{|\Omega|}{\pi^2\sqrt{m^2-1}}\int \frac{dy}{\sqrt{(1-y^2)(y_1-y)(y-y_2)}}.
\end{align}
Here the boundaries of integration are determined by Eq.~(\ref{ab}), which reads $y_1\ge y \ge y_2$, implying $|y|\leq 1$.
There are two cases, in which these conditions can be satisfied by the integration variable  $y$: 
\begin{align}\label{con1}
      |y_1|\leq 1, |y_2|>1,\notag\\
      |y_1|>1, |y_2|\leq 1.
\end{align}
First one leads to the integration region $y_1\le y\le 1$, and the second one to $-1\le y\le y_1$. 
To perform integration we use the identity (3.147.5) in \cite{Gradshteyn}, namely
\begin{align}
\int_{u_2}^{u_3}\frac{dy}{\sqrt{(u_4-y)(u_3-y)(y-u_2)(y-u_1)}}\notag\\
=\frac{2}{\sqrt{(u_4-u_2)(u_3-u_1)}}K\left[ \sqrt{\frac{(u_3-u_2)(u_4-u_1)}{(u_4-u_2)(u_3-u_1)}}\right],
\end{align}
where $u_4>u_3>u_2>u_1$. The first line of Eq.~(\ref{con1}) corresponds to $u_1=-1$, $u_2=y_1$, $u_3=1$ and $u_4=y_2$ and the second line corresponds to $u_1=y_1$, $u_2=-1$, $u_3=y_2$ and $u_4=1$. The results of the integration in both cases are identical
\begin{align}\label{r1yf}
\rho(\Omega)=\frac{\sqrt{2}}{\pi^2} \frac{|\Omega|}{\sqrt{m^2-1}}\frac{1}{\sqrt{y_2-y_1}}K\left[ \sqrt{\frac{(1-y_1)(1+y_2)}{2(y_2-y_1)}}\right].
\end{align}

The last step of this evaluation is to express the obtained formula in terms of $m$ and $\Omega$. As follows from Eq.~(\ref{yy}) for all $\Omega^2>2-m^2$ we can set $y_1=\tilde{y}_1$, $y_2=\tilde{y}_2$, where $\tilde{y}_1$, $\tilde{y}_2$ are given by Eq.~(\ref{y12}).
Then the solution of the inequalities (\ref{con1}) is given by the condition $(|m|-2)^2<\Omega^2<(|m|+2)^2$. After making sure that $2-m^2<(|m|-2)^2$ we substitute the same $y_1$, $y_2$ into Eq.~(\ref{r1yf}) to obtain
\begin{align}
\rho_{\text{I}}(\Omega) =\frac{\sqrt{2}}{\pi^2}\frac{|\Omega|}{\sqrt{\Omega^2+m^2-2}} \cdot \nonumber\\
\cdot K\left[ \sqrt{\frac{[\Omega^2-(m-2)^2][(m+2)^2-\Omega^2]}{8(\Omega^2+m^2-2)}}\right]  ,
\end{align}
where we denote this result by $\rho_{\text{I}}(\Omega)$. 

The final expression for the DOS for the case $|m|>1$ can be presented as
\begin{equation}\label{rhog11}
\rho_{|m|>1} (\Omega) =  \begin{cases}
   \rho_{\text{I}} (\Omega) ,\;\;\;\;\text{if}\;\; (|m|-2)^2\leq\Omega^2\leq(|m|+2)^2,\\
   0, \;\;\;\;\;\;\;\;\;\;\;\text{otherwise}.
 \end{cases}
\end{equation}
Note, that this expression reduces to the case $m^2=1$.  In this case $|m|=1$ Eq.~(\ref{rhog11}) coincides exactly with Eq.~(\ref{r1}). 

\subsubsection{Case $|m|<1$}

We repeat all the reasoning of the previous subsection for the case $|m|<1$. In this case the factor $|m|-1$ is negative, and the nonzero DOS obtained from Eqs.~(\ref{Iy}) and (\ref{rho0}) in terms of the variable $y$ is of the form
\begin{align}\label{ry}
\rho(\Omega)=\frac{|\Omega|}{\pi^2\sqrt{1-m^2}}\int \frac{dy}{\sqrt{(1-y^2)(y-y_1)(y-y_2)}},
\end{align}
 where $|y|\leq 1$ by the definition. The boundaries of the integration are determined by Eq.~(\ref{ab}), that gives $y\le y_1$ or $y\ge y_2$. There are three different possibilities for $y$ to satisfy these conditions,
\begin{align}\label{con2}
      |y_1|\leq 1, |y_2|>1, \notag\\ 
      |y_1|>1, |y_2|\leq 1,\notag \\ 
      |y_1|<1, |y_2|< 1.
\end{align}
We are considering each of them.

To perform integration in Eq.~(~\ref{ry}) we use the identities (3.147.3) and (3.147.7) in \cite{Gradshteyn} respectively, that for the case of complete elliptical integrals have the same right hand side, namely
\begin{subequations}
\begin{align}
\int_{u_1}^{u_2}\frac{dy}{\sqrt{(u_4-y)(u_3-y)(u_2-y)(y-u_1)}} = \label{Gr3}\\
=\int_{u_3}^{u_4}\frac{dy}{\sqrt{(u_4-y)(y-u_3)(y-u_2)(y-u_1)}} = \label{Gr7}\\
=\frac{2}{\sqrt{(u_4-u_2)(u_3-u_1)}}K\left[ \sqrt{\frac{(u_4-u_3)(u_2-u_1)}{(u_4-u_2)(u_3-u_1)}}\right],\notag
\end{align}
\end{subequations}
where $u_4>u_3>u_2>u_1$. 

The first line of Eq.~(\ref{con2}) gives the boundaries of integration as $-1\le y\le y_1$ in the integral (\ref{ry}). The integration can be performed by using Eq.~(\ref{Gr3}) with $u_1=-1$, $u_2=y_1$, $u_3=1$ and $u_4=y_2$.  
The second line of Eq.~(\ref{con2}) gives the boundaries of integration as $y_2\le y\le 1$ in the integral (\ref{ry}). The integration can be performed by using Eq.~(\ref{Gr7}) with $u_1=y_1$, $u_2=-1$, $u_3=y_2$ and $u_4=1$. The results in these two cases are identical   
\begin{align}\label{r12}
\rho(\Omega)=\frac{\sqrt{2}}{\pi^2} \frac{|\Omega|}{\sqrt{1-m^2}}\frac{1}{\sqrt{y_2-y_1}}K\left[ \sqrt{\frac{(1+y_1)(y_2-1)}{2(y_2-y_1)}}\right] .
\end{align}

The third line of Eq.~(\ref{con2}) splits integral (\ref{ry}) into the sum of two integrals with boundaries $-1\le y\le y_1$ and $y_2\le y\le 1$. The first one is performed by using Eq.~(\ref{Gr3}) and the second one is performed by using Eq.~(\ref{Gr7}), where we put $u_1=-1$, $u_2=y_1$, $u_3=y_2$ and $u_4=1$ for both integrals. The results of integrations are identical and summation is reduces to multiplication by $2$. The final expression reads
\begin{align}\label{r3}
\rho(\Omega)=\frac{2\sqrt{2}}{\pi^2}  \frac{|\Omega|}{\sqrt{1-m^2}}\frac{1}{\sqrt{(1-y_1)(1+y_2)}} \cdot \notag\\
 \cdot K\left[ \sqrt{\frac{(1+y_1)(1-y_2)}{(1-y_1)(1+y_2)}}\right] .
\end{align}

As the last step we are expressing the formulas (\ref{r12}) and (\ref{r3}) in terms of $m$ and $\Omega$. Eq.~(\ref{r12}) corresponds to the first two lines of Eq.~(\ref{con2}). 
As follows from Eq.~(\ref{yy}) for all $\Omega^2>2-m^2$ we can set $y_1=\tilde{y}_2$, $y_2=\tilde{y}_1$, where $\tilde{y}_1$, $\tilde{y}_2$ are given by Eq.~(\ref{y12}).
Note that the first two lines of Eq.~(\ref{con2}) are equivalent to Eq.~(\ref{con1}) and are symmetrical with respect to permutation of $y_1$ and $y_2$. So we can repeat the same reasoning as for the case $|m|>1$. The solution of this condition with respect to $\Omega^2$ gives the same region $(|m|-2)^2<\Omega^2<(|m|+2)^2$ and substitution of $y_1$ and $y_2$ into Eq.~(\ref{r12}) results in the same DOS $\rho_{\text{I}}(\Omega)$, given by Eq.~(\ref{rhog11}).

To evaluate Eq.~(\ref{r3}) corresponding to the third line of Eq.~(\ref{con2}) we notice that this line is symmetric with respect to permutation of $y_1$ and $y_2$. Rewriting it in terms of $\Omega^2$ we obtain the region $m^2>\Omega^2>(|m|-2)^2$ in which Eq.~(\ref{r3}) is valid. This region is split into two parts by Eq.~(\ref{yy}). In particular, for $2-m^2>\Omega^2>(|m|-2)^2$ we set $y_1=\tilde{y}_2$, $y_2=\tilde{y}_1$, and substitute them to Eq.~(\ref{r3}) to obtain 
\begin{align}
\rho_{\text{II}}(\Omega) =\frac{8}{\pi^2} \frac{|\Omega|}{|\Omega^2+m^2|}\cdot \nonumber \\
\cdot K\left[ \sqrt{\frac{[\Omega^2-(m-2)^2][\Omega^2-(m+2)^2]}{(\Omega^2-m^2)^2}}\right],
\end{align}
where we denote this result by $\rho_{\text{II}}(\Omega)$. 
In the region $m^2>\Omega^2>2-m^2$ we set $y_1=\tilde{y}_1$, $y_2=\tilde{y}_2$, and substitution  to Eq.~(\ref{r3}) gives
\begin{align}
\rho_{\text{III}}(\Omega) =\frac{8}{\pi^2}\frac{|\Omega|}{\sqrt{[\Omega^2-(m-2)^2][\Omega^2-(m+2)^2]}} \cdot \nonumber \\
\cdot K\left[\sqrt{\frac{(\Omega^2-m^2)^2}{[\Omega^2-(m-2)^2][\Omega^2-(m+2)^2]}}\right],
\end{align}
where we denote this result by $\rho_{\text{III}}(\Omega)$. 
 The final expression for the DOS for the case $|m|<1$ can be presented as
\begin{equation}\label{rhog1}
\rho_{|m|<1} (\Omega) =  \begin{cases}
   \rho_{\text{I}}(\Omega) ,\;\;\;\;\text{if}\;\; (|m|-2)^2<\Omega^2\leq (|m|+2)^2,\\
\rho_{\text{II}}(\Omega) ,\;\;\;\text{if}\;\; 2-m^2\leq\Omega^2\leq(|m|-2)^2,\\
\rho_{\text{III}}(\Omega) ,\;\;\text{if}\;\; m^2\leq\Omega^2< 2-m^2,\\
   0, \;\; \;\;\;\;\;\;\;\;\;\text{otherwise}.
 \end{cases}
\end{equation}
Note, that the DOS for all values of parameter $m$ does not depend on the sign of $m$ and, therefore is symmetric with respect to $m$ and $-m$ in contrast to the dispersion relations.

\section{Summary of analytical results and plots of total DOS}

\begin{figure}[h!]
\centering 
{\includegraphics[width=0.5\textwidth]{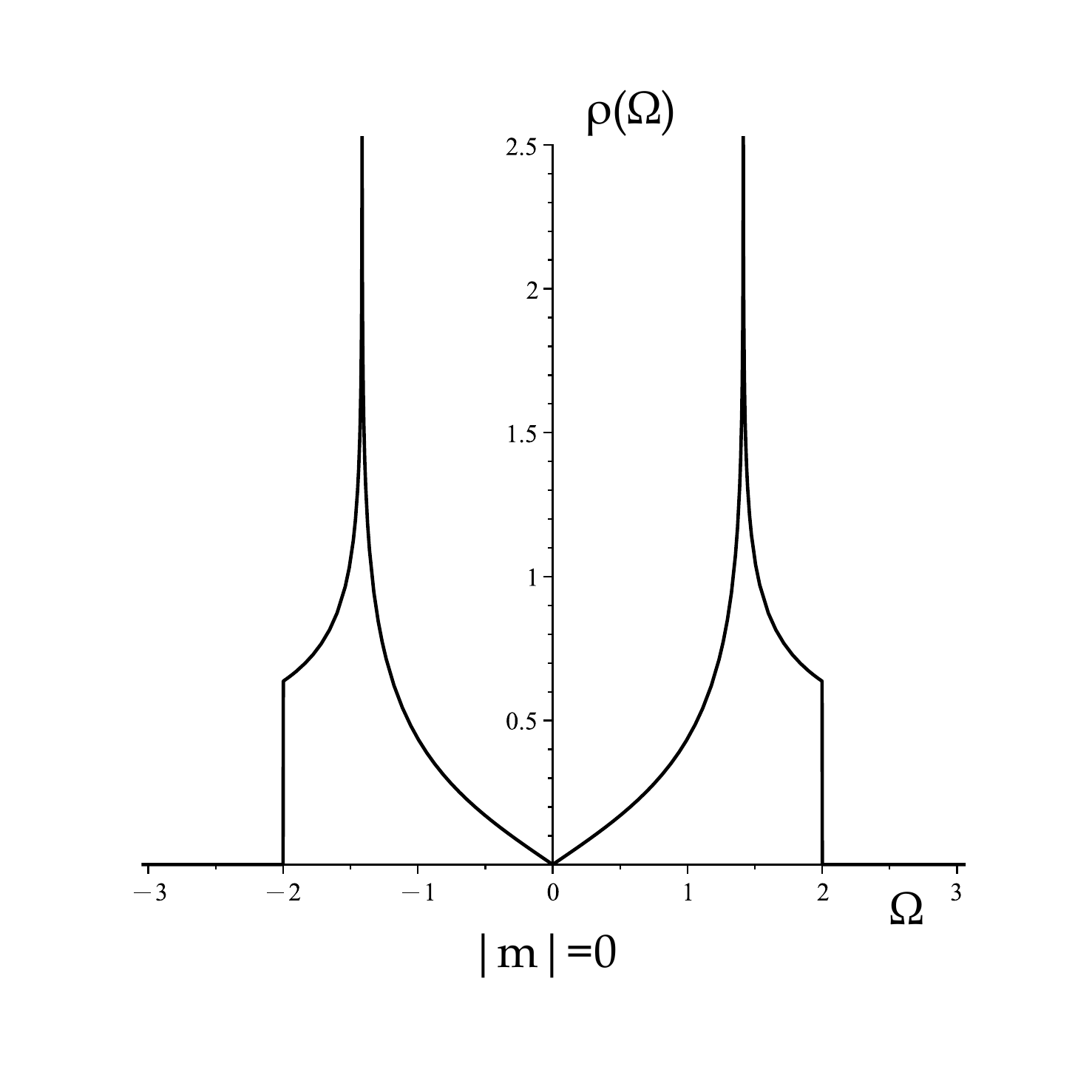}}
\caption{Total density of states of the QWZ model at $m=0$. }
\label{dos0}
\end{figure}

\begin{figure}[h!]
\centering
{\includegraphics[width=0.5\textwidth]{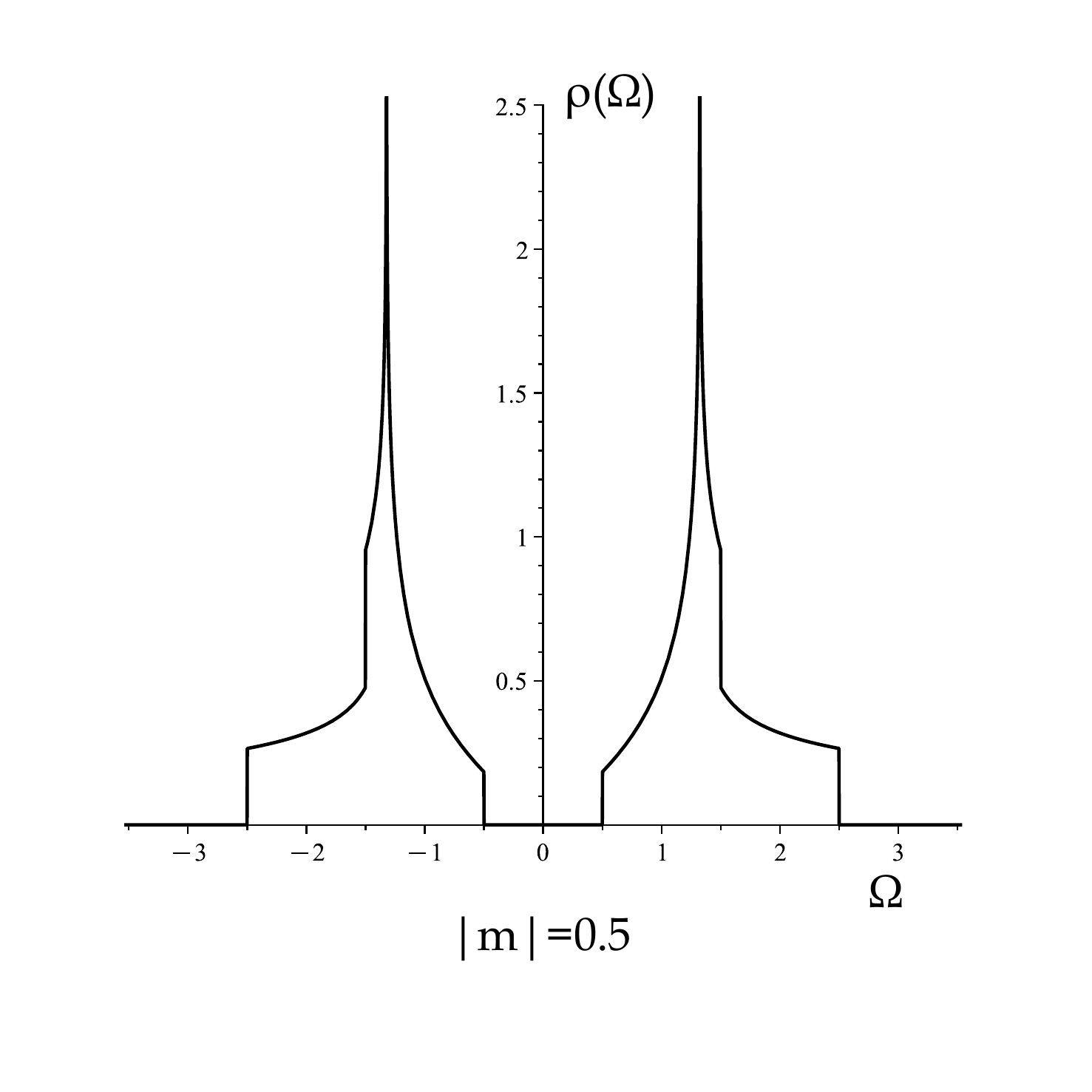}}
\caption{Total density of states of the QWZ model at $m=\pm 0.5$.}
\label{dos05}
\end{figure}

\begin{figure}[h!]
\includegraphics[width=0.5\textwidth]{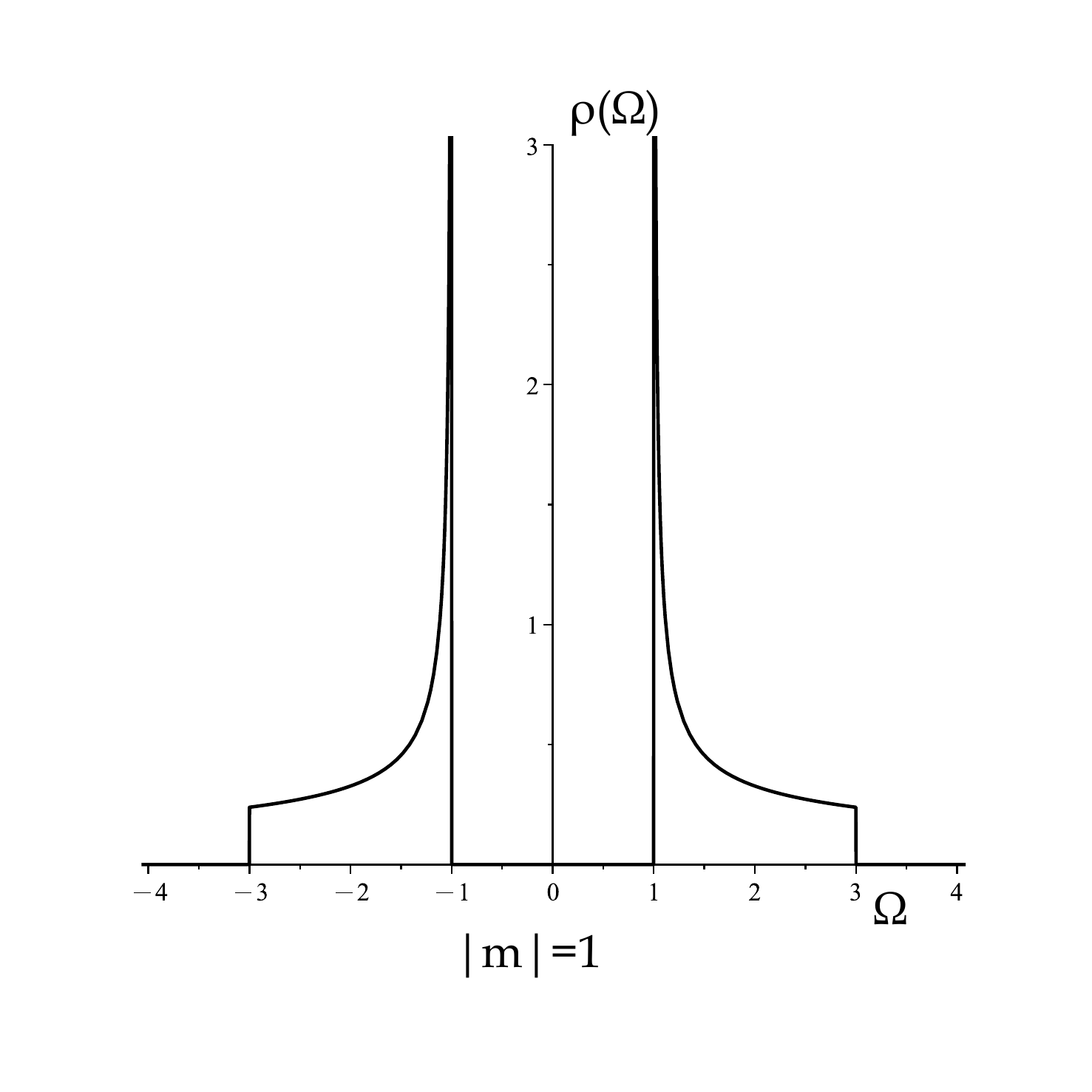}
\caption{Total density of states of the QWZ model at $m=\pm 1$.}
\label{dos1}
\end{figure}

\begin{figure}[h!]
\includegraphics[width=0.5\textwidth]{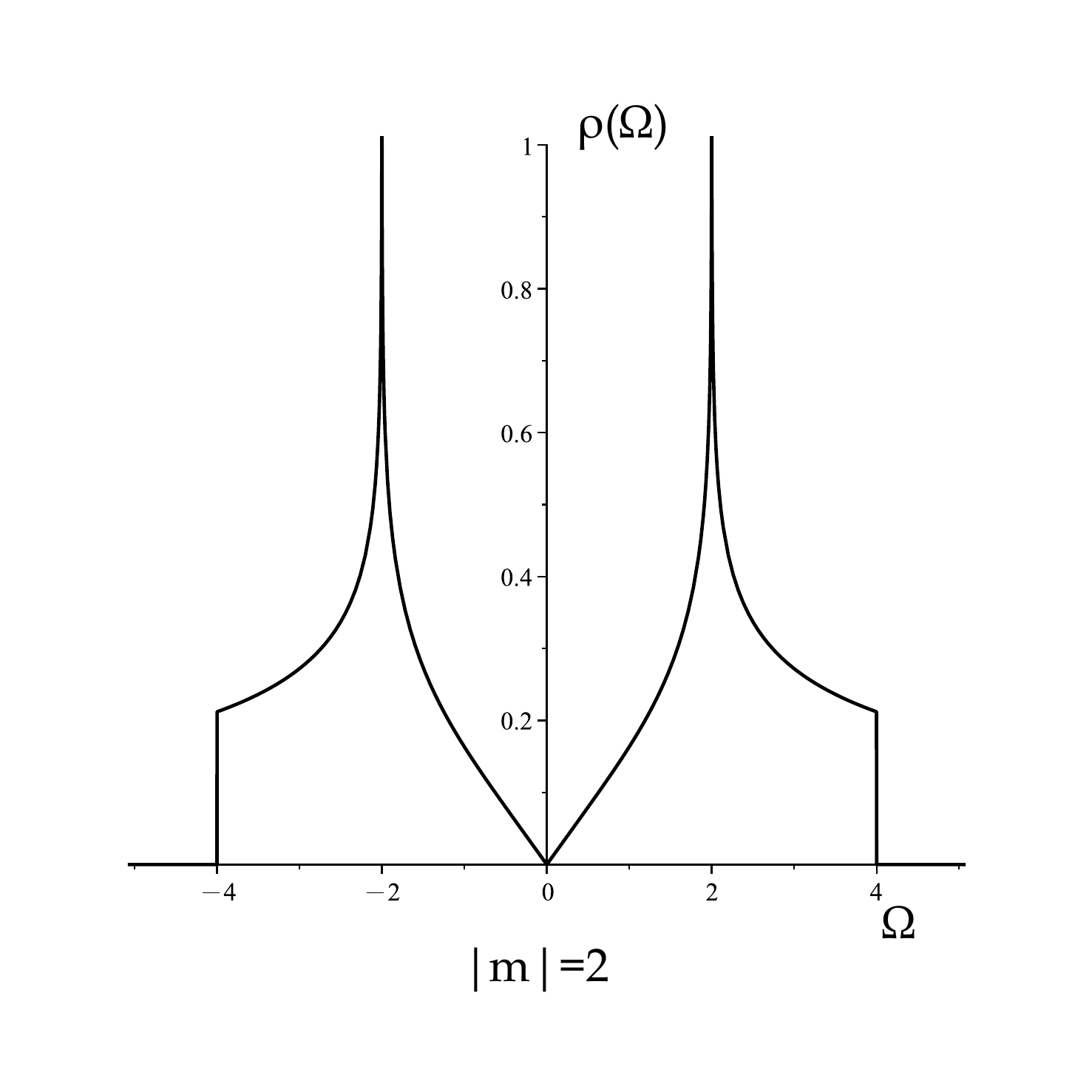}
\caption{Total density of states of the QWZ model at $m=\pm 2$.}
\label{dos2}
\end{figure}

\begin{figure}[h!]
\includegraphics[width=0.5\textwidth]{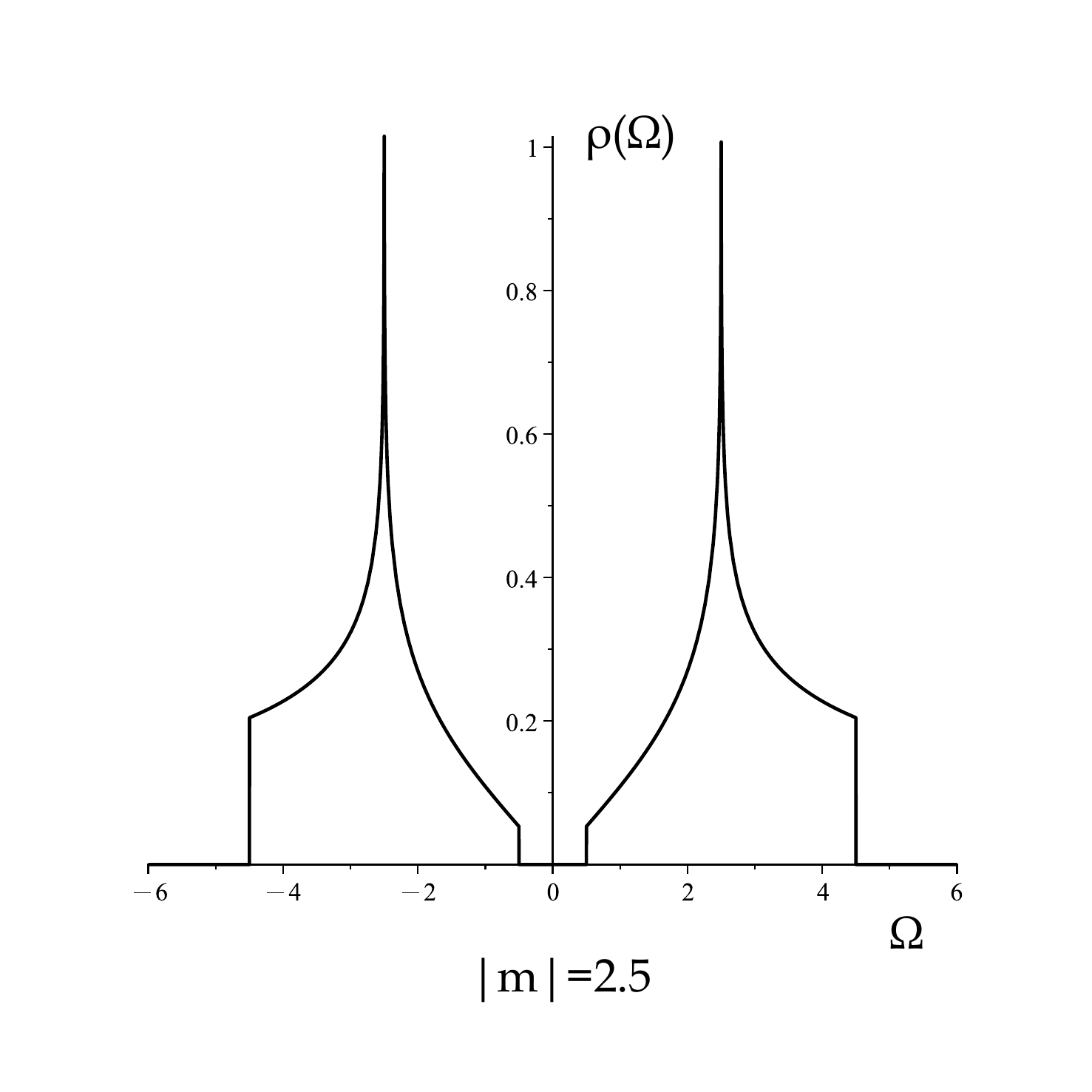}
\caption{Total density of states of the QWZ model at $m=\pm 2.5$.}
\label{dos25}
\end{figure}

In this Section we present plots of the total DOS for different $m$, corresponding to the dispersion relations shown  in Section II,  and we provide consistent  analyzes   of them. The plots are given in Figs.~\ref{dos0}-\ref{dos25}, that appear in the same order as the dispersion relations in Figs.~\ref{disp0}-\ref{disp25}.  In these DOS  plots the ranges of the vertical axis are different but all DOS are normalized to two, corresponding to two bands. We see that the shapes of the DOS for the QWZ model are much richer and with more additional features as compared, for example,  to the hexagonal lattice with nearest neighbor hopping.  Apart of the symmetry $m$ and $-m$, mentioned  above,   the total DOS is symmetric with respect to $\Omega$ and $-\Omega$. It is clearly visible in analytic formulae of the DOS, Eqs.~(\ref{r1}, \ref{rhog11}, \ref{rhog1}), where $\Omega$ is only present as either $\Omega^2$ or $|\Omega|$. 

The typical shape of the DOS for $0<|m|<1$ is shown in Fig.~\ref{dos05}. We can see that each nonvanishing part of the plot has three sections, in each of the section the function is described by $\rho_\text{I}$, $\rho_\text{II}$, or $\rho_\text{III}$, cf Eq.~(\ref{rhog1}). Specifically, in the plot in Fig.~\ref{dos05} for $|m|=0.5$, these sections are separated by the points $|\Omega|=\sqrt{1.75}\approx1.32$ and $|\Omega|=1.5$.  At the energies $\pm\sqrt{2-m^2}$, separating two nearby sections, the DOS has infinite peaks. 

Another characteristic feature of the system with $0<|m|<1$ is the opening of the band gap of the width $2|m|$. It can be seen in the corresponding dispersion relations, cf. Fig.~\ref{disp05}, and in the plots of DOS, where $\rho(\Omega)=0$ in the range $-|m|<\Omega<|m|$. At the half-filling such system is a topological insulator \cite{QWZ,Short}.

The  gap is closed for $m=0$ as seen in Fig.~\ref{dos0}, when 
 the DOS has a pseudogap at $\Omega=0$ (DOS vanishes at a single point). It corresponds to formation of the Dirac cones at $X=(\pm\pi,0)$ and $Y=(0,\pm\pi)$ points in the Brillouin zone, that is easy to see in the plots of the dispersion relations, cf. Fig.~\ref{disp0}. At the half-filling such system is a semi-metal. 

The special case $|m|=1$, given by Eq.~(\ref{r1}), is shown in Fig.~\ref{dos1} and the corresponding dispersion relations are in Fig.~\ref{disp1}. Flat parts in the dispersion relations, i.e. lines along which\textcolor{black}{the gradient of the dispersion} ${\bf \nabla}_{\bf k}\epsilon_{\pm}({\bf k})=0$, give rise to the appearance of sharp peaks in the DOS. Despite of the fact that these flat parts are different for $m=1$ and $m=-1$, the shape of the DOS is the same. In this $|m|=1$ case  the system is a topological insulator at the half-filling \cite{QWZ,Short}.
 
The typical shape of the DOS for $|m|>1$ (without $|m|=2$) is shown in Fig.~\ref{dos25}. The nonzero part of the DOS function is described by the single elliptic integral, cf. Eq.~(\ref{rhog11}). It has two symmetrical infinite peaks at energies $\pm|m|$. The system has the band gap of the width $2||m|-2|$ in the dispersion relation, shown in Fig.~\ref{disp25}. In the plot of the DOS the gap corresponds to $\rho(\Omega)=0$ in the range $-||m|-2|<\Omega<||m|-2|$. At the half-filling the system with $|m|<2$ is a  topological insulator and the system with $|m|>2$ is  a  trivial insulator \cite{QWZ,Short}.  

For $|m|=2$ the band gap closes and a pseudogap appears as it is shown in the plot of the DOS in Fig.~\ref{dos2}. Formation of the Dirac cones can be seen at the corresponding dispersion relations in Fig.~\ref{disp2}. For $m=2$ the gap is closed at $M=(\pm\pi,\pm\pi)$ and $(\pm\pi,\mp\pi)$ points in the Brillouin zone and for $m=-2$ it is closed at $\Gamma=(0,0)$ point. At the half filling such system is a semi-metal. 

\textcolor{black}{Having the exact, analytic expressions for the DOS in terms of the elliptic integrals we can prove  rigorously that the singularities, present for all $m$ values, are of the logarithmic  divergence type, very similarly as in case of the square lattice. It comes form the exact properties of the elliptic integrals \cite{abramowitz}. In numerical approaches to the DOS such  conclusion would be hard to achieve rigorously since the DOS would typically depend on an artificial broadening parameter or an arbitrary truncation of an infinite set of recursive equations. }

\section{Subtle features seen in total DOS} 

In this Section, we discus additional subtle features and general trends that can be observed for the DOS of the considered QWZ model. 

\subsection{Additional finite peaks}

\begin{figure}[h!]
\includegraphics[width=0.4\textwidth]{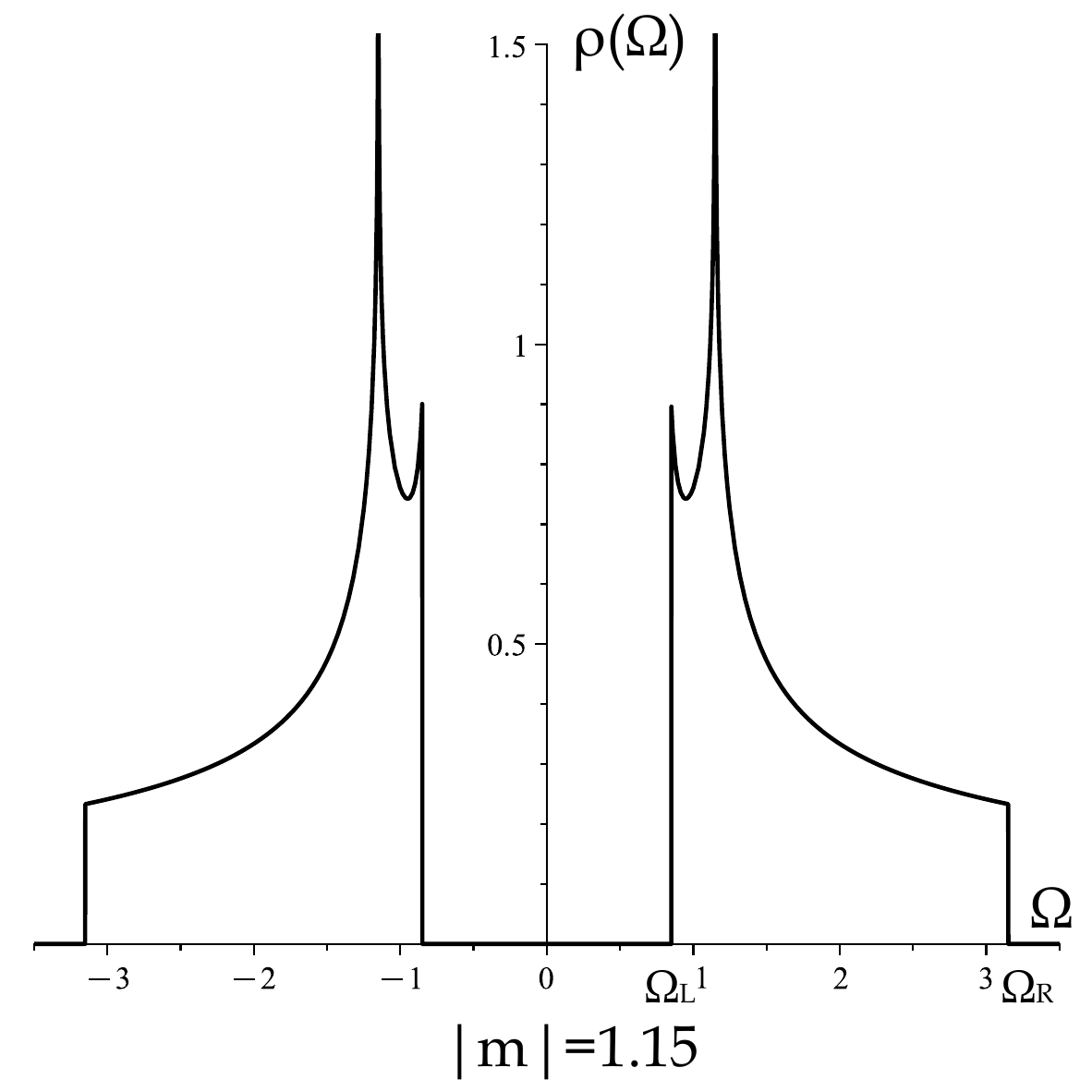}
\caption{Total density of states of the QWZ model at $m=\pm 1.15$.}
\label{dos115}
\end{figure}

As discussed earlier, the  total DOS is symmetric with respect to $m$ and $-m$ and as well as it is symmetric with respect to $\Omega$ and $-\Omega$. It has two infinite peaks located at $\pm\Omega_\infty$, where the elliptical integral $K(1)=\infty$, i.e., 
\begin{equation}\label{}
\Omega_\infty= \begin{cases}
  \sqrt{2-m^2} ,\;\;\;\text{if}\;\;|m|<1,\\
{|m|},\;\;\;\;\;\;\;\;\;\;\;\;\text{if}\;\;|m|\geq 1.
 \end{cases}
\end{equation}
We find, that for the values of $|m|$ slightly larger than $1$, two additional finite peaks appear at the edges of a band gap given by $\pm\Omega_L$, for example,  it is shown in Fig.~\ref{dos115} for the case of $|m|=1.15$. At larger $|m|$ these peaks disappear.

\subsection{Widths of the band gap and the bands} 

Let $\Delta$ be the width of the band gap. Its  value for arbitrary $m$ (in units with $t=1$) is given by the   simple expression
\begin{equation}\label{}
\Delta= 2\begin{cases}
  {|m|} ,\;\;\;\;\;\;\;\;\;\;\;\text{if}\;\;|m|<1,\\
{||m|-2|},\;\;\;\text{if}\;\;|m|\geq 1.
 \end{cases}
\end{equation}
Edges of the band gap are located at energies $\pm\Omega_L$, where $\Omega_L=\Delta/2$. 

On the other hand, the upper and lower bounds of the energy spectrum (the dispersion relations $\epsilon_{\pm}({\bf k})$) are at  $\pm\Omega_R$, where $\Omega_R=|m|+2$. The width of each of the two bands $\epsilon_{\pm}({\bf k})$ is defined as $W=\Omega_R-\Omega_L$, since  $\Omega_R>\Omega_L$. Its  value is explicitly given by  
\begin{equation}\label{}
W= 2\begin{cases}
 1 ,\;\;\;\;\;\;\;\;\;\;\;\text{if}\;\;|m|<1,\\
|m|,\;\;\;\;\;\;\;\text{if}\;\;1\geq|m|\geq 2,\\
 2 ,\;\;\;\;\;\;\;\;\;\;\;\text{if}\;\;|m|>2.
 \end{cases}
\end{equation}

\begin{figure}[h!]
\includegraphics[width=0.4\textwidth]{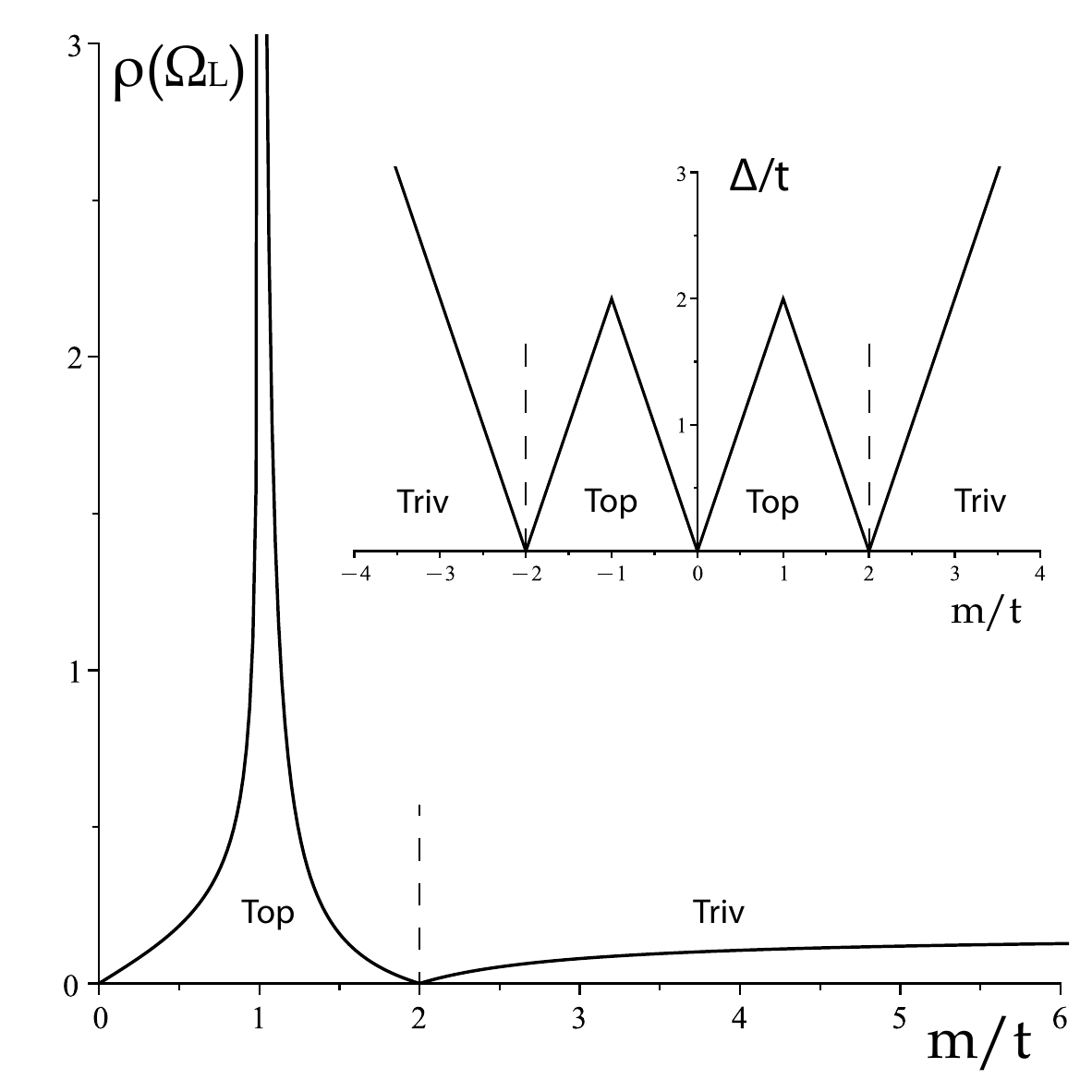}
\caption{The value of the DOS at the edges of a band gap, i.e. at $\Omega=\Omega_L$, vs. the parameter $m$. The Inset shows a dependence of the gap width vs. $m$. The topological and trivial insulators are indicated in the figures.}
\label{dos_edge}
\end{figure}

The dependence of the gap width $\Delta$   as a function of  $m$ is shown in the Inset of Fig.~\ref{dos_edge}. It is seen that when  $\Delta<2$ the gap of the same width is opened for the six different values of the parameter $m$, at half-filling four of these $m$  will correspond to topological insulators  and other two $m$ will correspond to the trivial insulators.

\begin{figure}[h!]
\includegraphics[width=0.4\textwidth]{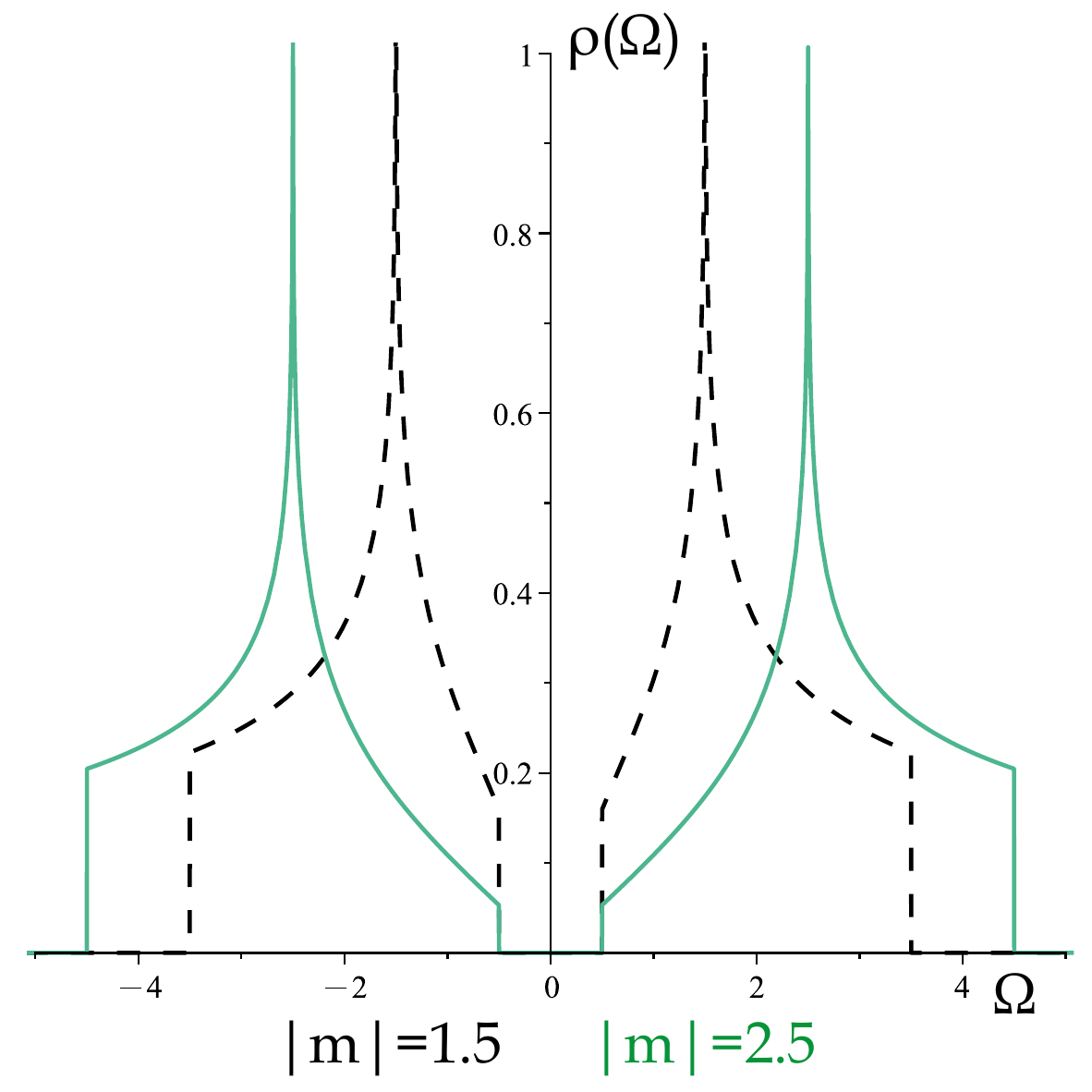}
\caption{Comparison of the two DOS of the QWZ model, which are characterized by the same gap. Total density of states at $m=\pm 1.5$ (black dashed line) at half-filling corresponds to topological insulator. Total density of states at $m=\pm 2.5$ (green solid line) at half-filling corresponds to trivial insulator. Enhancement of the DOS for topological case is seen.}
\label{dos_comp}
\end{figure}

To discuss further  we chose the case with the width $\Delta=1$, which  is possible for $|m|=0.5$, $1.5$, and $2.5$. The DOS for $|m|=1.5$ and $2.5$ are shown in Fig.~\ref{dos_comp} whereas  the DOS for $|m|=0.5$ is in Fig.~\ref{dos05}. Interestingly, it can be seen that the DOS at the edges of the  band gap, i.e., $\rho(\Omega=\pm \Omega_L)$, for the topological insulator ($|m|=0.5$ and $1.5$) is  larger than the one for the trivial case $|m|=2.5$.\textcolor{black}{ Such a behavior can be explained by the fact that the topological phase with $|m|<2$ is associated  with the overlap of the bands and the so-called band inversion phenomenon. If the off-diagonal terms $h_x({\bf k})$ and  $h_y({\bf k})$ were neglected the bands only overlap, resulting  enhancement of the DOS in the overlapping regime of energies. We note however, that enhancing of the DOS due to this mechanism is not a sufficient condition to indicate the system as  topologically non-trivial. On the other hand, a band inversion is a necessary condition for that, at least in a broad class of tight-binding models.} The DOS at the edges of the band gap can be obtained analytically using the properties of the elliptical integrals $K(0)=\pi/2$, namely
\begin{equation}\label{}
\rho (\Omega_L) = \frac{1}{2\pi} \begin{cases}
  {2|m|}/{\sqrt{1-m^2}} ,\;\;\;\;\;\;\;\;\;\text{if}\;\;|m|<1,\\
{||m|-2|}/{({|m|-1})} ,\;\;\;\text{if}\;\;|m|\geq 1.
 \end{cases}
\end{equation}
The dependence of $\rho (\Omega_L)$ as a function of  $m$ is shown in Fig.~\ref{dos_edge}. For $|m|>2$ it is a finite smooth  function. When  $|m|<2$ the function is singular and diverges at $m=\pm1$. 

\section{Spectral moments of total DOS} 

In this Section we present  results on the spectral  moments of the total DOS. The general formula for the spectral moment of the order $n$ reads
\begin{equation}\label{Mn}
M_n=\int{\Omega^n\rho(\Omega)d\Omega}.
\end{equation}
Integrals of the form (\ref{Mn}) can be calculated analytically on the base of our analytic results from section III. Some of the calculations are shown in Appendix \ref{app} to demonstrate details of the integration technique.

The moment of the zero order represents normalization integral $M_0=2$.
All moments of the odd orders are zero because the DOS is an even function
of $\Omega$.

The moments of the even orders $n$ are found as the polynomials in the parameter $m$ of the same orders:
\begin{align}\label{pol}
M_2=2(m^2+2),\notag\\
M_4=2(m^4+8m^2+5),\notag\\
M_6=2(m^6+18m^4+51m^2+14),\notag\\
M_8=2(m^8+32m^6+210m^4+284m^2+42). 
\end{align}
This sequence can be continued if needed. Note, that all moments have a factor of $2$, which corresponds to two symmetrical bands.

\section{Conclusions and outlooks} 

In this paper we derived the analytic formulae of the DOS of the QWZ Hamiltonian, a generic model for\textcolor{black}{Chern} topological insulators in two dimensions. The results are expressed in terms of the complete elliptic integrals. Analytic expressions for the DOS are rare in general. Our results extend the class of\textcolor{black}{tight-binding} models where the\textcolor{black}{exact,} analytic DOS is known. 

We discussed in details the plots of the DOS and compared them with the dispersion relations for the same value of the parameter $m$. Some additional finite peaks in the DOS were identified. We provided explicit formulae for the gap width and for the width of the bands in the QWZ model. We also found that for the same gap width the topological system has larger DOS at the gap edge as compared with the trivial case. Apparently, in the topological case a more spectral weight is redistributed close to the band gap.\textcolor{black}{It is due to the bands overlap and the inversion band phenomenon.} Finally, we obtained expressions of the spectral moments of the QWZ model. They are  polynomials of the parameter $m$, controlling the topology, and are  of the same  orders as the orders of the corresponding moments.\textcolor{black}{On the base of the analytic results we can identify exactly the position of van Hove singularities and their logarithmic type. }

The analytic DOS will be useful in determining  thermodynamics\textcolor{black}{(specific heat, compressibility, or magnetic susceptibility)} or\textcolor{black}{linear} response\textcolor{black}{(dc  conductivity)} of the QWZ Chern topological insulator. It will also simplify the dynamical mean-field theory study of the QWZ model when a\textcolor{black}{local} Hubbard type of the interaction is added to the Hamiltonian, cf. \cite{dmft1,dmft2,dmft3}.\textcolor{black}{A final remark, as mathematicians say: "although a donut and a cup are topologically equivalent, one cannot drink a coffee from a donut. The shape, i.e., the geometry matters". Similarly,  all QWZ model's physical properties, not only topological ones, are importnt to be known in general.} We\textcolor{black}{also} hope that by using similar methods other DOS  can be obtained in analytic form, for example for  the Haldane model\textcolor{black}{of topological insulator} on a hexagonal lattice. 

\section{Acknowledgment}

We thank for the financial support of the {\it Excellence Initiative - Research University} (IDUB) via the grant under the program New Ideas - Ukraine. K.B.  also acknowledges the support of the  Deutsche Forschungsgemeinschaft under the {\it Transregional Collaborative Research Center} TRR360.

\appendix
\section{\textcolor{black}{Complete} elliptic integrals of the first and second kind \label{def}}

Expressions for\textcolor{black}{complete }elliptic integrals, which we are using, are the following
 \begin{eqnarray}\label{KK}
K(q)=\int_0^{\pi/2}\frac{d\beta}{\sqrt{1-q^2\sin^2\beta}},
\end{eqnarray}
 \begin{eqnarray}\label{}
E(q)=\int_0^{\pi/2}{d\beta}{\sqrt{1-q^2\sin^2\beta}} ,
\end{eqnarray}
where\textcolor{black}{ $q$ is the modulus of the elliptic integral and $q'$ is the complementary modulus, i.e., $q'^2=1-q^2$. The elliptic integrals can also be expressed as sums or series of rational functions, \cite{Gradshteyn}.}
\textcolor{black}{The above definitions are traditionally used by mathematicians. Note, that in Python libraries $q^2$ is substituted as an argument of $K$.}

\section{Calculation of spectral moments in details\label{app}}

Here we show how to calculate some of the integrals (\ref{Mn}). Though the spectral moment of any order can be easily calculated numerically, the analytical derivation is of special value and methodological interest.

We start with the simple case of $|m|=1$. We express $q^2={9-\Omega^2}/{8}$ and $q'^2=(\Omega^2-1)/8$. For all values of $\Omega^2$ from the region where $\rho(\Omega)$ is nonzero the modulus is $0\leq q\leq 1$. In these terms the integral, normalizing the DOS, takes the form 
\begin{equation}\label{M0}
\frac{M_0}{2}=\frac{4}{\pi^2}\int_0^1{K(q)dq'}.
\end{equation}
Using the identity (6.141.2) in \cite{Gradshteyn},
\begin{equation}\label{}
\int_0^1{K(q')dq}=\frac{\pi^2}{4}.
\end{equation}
gives the correct value of normalization. 

The second moment takes  the form 
\begin{equation}\label{m2}
\frac{M_2}{2}=\frac{4}{\pi^2}\int_0^1(9-8q^2){K(q)dq'}.
\end{equation}
It can be rewritten as $M_2/2=M_0/2+32J/\pi^2$, where
\begin{equation}\label{J}
J=-\int_0^1q'K(q)qdq.
\end{equation}
Here we used $q'dq'=-qdq$. The integral $J$ can be determined by parts using the following substitution $dv=K(q)qdq$ and $u=q'$. It allows us to use the indefinite integral (5.112.3) in \cite{Gradshteyn}, namely
\begin{align}\label{}
v=\int K(q)qdq=E(q)-q'^2K(q),
\end{align}
where $E(q)$ is the elliptic integral of the second kind. Integration by parts gives $J=\int_0^1 E(q)dq'-J$, where it is used that $E(0)=K(0)=\pi/2$ and $E(1)=1$.
Using the identity (6.148.2) in \cite{Gradshteyn}
\begin{equation}\label{}
\int_0^1{E(q')dq}=\frac{\pi^2}{8}
\end{equation}
we finally obtain $J={\pi^2}/{16}$ and $M_2=6$. 

The moment of the order $n$ is expressed as a combination of integrals of the product of an elliptic function and an even power of the modulus,
\begin{equation}\label{Mc}
{M_n}=\sum_{i=0}^{n/2}C_{i}\int_0^1q^{2i}{K(q)}dq'.
\end{equation}
Here $C_i$ are real numbers, compare with Eq.~(\ref{m2}). So the same procedure  with integration by parts can be used to obtain moment of arbitrary order.

For $|m|>1$ the DOS have the infinite peaks at $\pm\Omega_\infty$. Integral over $\Omega$ in Eq.~(\ref{Mn}) is split into two: from $\Omega_L$ to $\Omega_\infty$ and from $\Omega_\infty$ to $\Omega_R$. In these regions the complementary modulus $q'$ take values from $1$ to $0$, and from $0$ to $1$ respectively. Denoting $x=\Omega^2$ we can present the moment of the order $n$ in the form 
\begin{align}\label{spl}
{M_n}=\frac{\sqrt{2}}{\pi^2}\int_0^1dq'{K(q)}\left[\frac{x_2^{n/2}}{\sqrt{x_2+m^2-2}}\left(\frac{dx_{2}}{dq'}\right)-\right.\notag
\\
\left.\frac{x_1^{n/2}}{\sqrt{x_1+m^2-2}}\left(\frac{dx_{1}}{dq'}\right)\right].
\end{align}
The substitution $x=x_{1,2}(q)$ is not a single-valued function: $x_{1,2}=m^2+4q'^2\mp 4q'\sqrt{m^2-q^2}$, where the sign $"+"$ corresponds to the first region and $"-"$ to the second one.
The expression in the square brackets is transformed algebraically to a polynomial function $\sum_{i=0}^{n/2}C_{i}(m^2)q^{2i}$, so for $n=0$ it is equal to $4\sqrt{2}$ and for $n=2$ it equal to $4\sqrt{2}(m^2+8q'^2)$. Then moments are of the form (\ref{Mc}), where coefficients $C_i=C_i(m^2)$ are the functions of $m^2$. The result of calculations leads to the expressions (\ref{pol}).  

For the case $|m|<1$ two regions of integration need to be considered separately: $|m|\leq\Omega\leq 2-|m|$ and $2-|m|\leq\Omega\leq 2+|m|$. In each of these regions the substitution ${x}={x}({q})$ is not a single-valued, so  each integral splits into two ones, similar to the case with $|m|>1$.

In the first region we obtain the following: For $|m|\leq\Omega\leq \sqrt{2-m^2}$ the substitution is $x_1=m^2-4q(q-\sqrt{1-m^2q'^2})/q'^2$. The limits of integration $x_1(q'=1)=m^2$ and $\lim_{q'\rightarrow0}x_1=2-m^2$. For $\sqrt{2-m^2}\leq\Omega\leq 2-|m|$ the substitution is $x_2=m^2+4(1-\sqrt{q^2+m^2q'^2})/q'^2$ with the limits of integration $x_2(q'=1)=(2-|m|)^2$ and $\lim_{q'\rightarrow 0}x_2=2-m^2$. 

In the second region we have $x_{3,4}=m^2+4{q'}^2\mp 4q'\sqrt{m^2-q^2}$ valid for $2-|m|\leq\Omega\leq \Omega_M$ and $\Omega_M\leq\Omega\leq 2+|m|$, respectively. 
Here $\Omega_M$ is defined by the condition $d q/d \Omega =0$ and $q^2={(\Omega^2-(m-2)^2)((m+2)^2-\Omega^2)}/{(\Omega^2+m^2-2)}/8$. It can be proven that $q(\Omega_M)=|m|$. 

Thus the formula for the $M_n$ takes the form
\begin{align}\label{Mx}
{M_n}=\frac{8}{\pi^2}\int_0^1dq'{K(q)}\left[\frac{x_2^{n/2}}{{x_2+m^2}}\left(\frac{dx_{2}}{dq'}\right)-\right.\notag
\\
\left.\frac{x_1^{n/2}}{\sqrt{(x_1-(m-2)^2)(x_1-(m+2)^2)}}\left(\frac{dx_{1}}{dq'}\right)\right]+\notag\\
\frac{\sqrt{2}}{\pi^2}\int_0^{|m|}dq{K(q)}\left[\frac{x_3^{n/2}}{\sqrt{x_3+m^2-2}}\left(\frac{dx_{3}}{dq}\right)-\right.\notag
\\
\left.\frac{x_4^{n/2}}{\sqrt{x_4+m^2-2}}\left(\frac{dx_{4}}{dq}\right)\right].
\end{align}
It should be noted, that these integrals are  difficult to calculate analytically. 
For example, for $M_0$ the substitution $x=x(q)$ in Eq.~(\ref{Mx}) results in
\begin{align}\label{MM0}
({M_0}-1)\frac{\pi^2}{8}=\int_0^{|m|}dq{K(q)}\frac{q}{\sqrt{m^2-q^2}}+\notag\\
\int_0^1{dq}\frac{K(q)}{q'^2}\left[\frac{q}{\sqrt{m^2+q^2(1-m^2)}}-\frac{q}{\sqrt{(1-m)^2+m^2q^2)}}\right],
\end{align}
where we used the identity (6.144) in \cite{Gradshteyn}
\begin{align}\label{}
\int_0^1 K(q)\frac{1}{1+q}dq=\frac{\pi^2}{8}.
\end{align}
Knowing the fact that $M_0=2$, the expression (\ref{MM0}) turns into an interesting relation between integrals containing $K(q)$. 

In order to determine  these integrals, the integrands can be expanded into an infinite series in $q^2$ followed by term-by-term integration. But the easiest way to prove Eq.~(\ref{pol}) for the case $|m|<1$ is the numerical integration.


\begin{thebibliography}{99}
\bibitem{Kane} M. Z. Hasan and C. L. Kane, Rev. Mod. Phys. {\bf 82}, 3045 (2010).
\bibitem{Zhang} X.L. Qi and S.C. Zhang, Rev. Mod. Phys. {\bf 83}, 1057 (2011). 
\bibitem{Laughlin} R. B. Laughlin, Phys. Rev. B {\bf 23}, 5632 (1981). 
\bibitem{nature-mat} J. Wang and S.C. Zhang, Nat. Materials {\bf 16}, 1062 (2017).
\bibitem{cond-mat} B.Q. Lv, T. Qian, and H. Ding,  Rev. Mod. Phys. {\bf 93}, 025002 (2021). 
\bibitem{cold} N.R. Cooper, J. Dalibard, and I.B.  Spielman,  Rev. Mod. Phys. {\bf 91}, 015005  (2019).
\bibitem{photonics} L. Lu, J. D. Joannopoulos and M. Soljacic,  Nat. Photonics {\bf 8},  821  (2014).
\bibitem{PT} A.P. Ramirez and   B. Skinner, Physics Today {\bf 73} (9), 30 (2020).
\bibitem{SSH} W. P. Su, J. R. Schrieffer, and A. J. Heeger, Phys. Rev. Lett. {\bf 42}, 1698 (1979). 
\bibitem{Rice} M. J. Rice and E. J. Mele, Phys. Rev. Lett. {\bf 49}, 1455 (1982).
\bibitem{Haldane} F.D.M. Haldane, Phys. Rev. Lett. {\bf 61}, 2015 (1988). 
\bibitem{QWZ} X.L. Qi, Y.S. Wu, and S.C. Zang, Phys. Rev. B {\bf 74}, 085308 (2006). 
\bibitem{Short} J. K. Asboth, L. Oroszlany, and  A. Palyi, {\it A Short Course on Topological Insulators}, Lecture Notes in Physics (LNP, vol. 919) (Springer, Cham, 2016). 
\bibitem{Potthoff} D. Kr\"uger and M. Potthoff, Phys. Rev. Lett. {\bf 126}, 196401 (2021). 
\bibitem{Streda} P. Streda, J.  Phys. C: Solid State Physics 1{\bf 5}, L717 (1982).
\bibitem{abramowitz} M. Abramowitz and I.A. Stegun, {\it Handbook of mathematical functions with formulas, graphs, and mathematical tables}, pages  587-626 (Dover publication, inc., New York, 1972). 
\bibitem{Hobson} J.P. Hobson and W. A. Nierenberg, Phys. Rev. {\bf89}, 662 (1953).
\bibitem{Ovchynnikov} V.O. Ananyev and M. I. Ovchynnikov, Cond. Mat. Phys. {\bf20}, 4, 43705 (2017).
\bibitem{Kogan} E. Kogan and G. Gumbs, Graphene {\bf10}, 1 (2021).
\bibitem{Jelitto68} R.J. Jelitto,  J. Phys. Chem. Solids {\bf 30}, 609 (1969).
\bibitem{Sakaji02}  A. Sakaji, R.S. Hijjawi, N. Shawagfeh, and J.M. Khalifeh, International Journal of Theoretical Physics, {\bf 41}, 973 (2002). 
\bibitem{Hijjawi04}  R.S. Hijjawi, J.H. Asad, A. Sakaji, and J.M. Khalifeh, International Journal of Theoretical Physics, {\bf 43}, 2299 (2004). 
\bibitem{Georges96} A. Georges, G. Kotliar, W. Krauth, and M.J. Rozenberg, Rev. Mod. Phys. {\bf 68}, 13 (1996). 
\bibitem{Ulmke98} M. Ulmke, Eur. Phys. J. B {\bf 1}, 301 (1998). 
\bibitem{Wahle98} J. Wahle, N. Bl\"umer, J. Schlipf, K. Held, and D. Vollhardt, Phys. Rev. B {\bf 58}, 12749 (1998).
\bibitem{Kapcia21} K.J. Kapcia, R. Lema\'nski,  and M.J. Zygmunt, J. Phys.: Condens. Matter {\bf 33},  065602 (2021). 
\bibitem{python} https://pypi.org/project/gftool/
\bibitem{ten-way} C.K. Chiu, J.C.Y. Teo, A.P. Schnyder, and S. Ryu, Rev. Mod. Phys. {\bf 88}, 035005  (2016).
\bibitem{footnote} As will be seen latter, when the gap is closed at $m=0$ and $|m|=2$ the bands touch each other at separated point(s) and the DOS is zero at the corresponding energy $\Omega=0$. Hence, the band resolved DOS is determined uniquely from the total DOS. 
\bibitem{Gradshteyn} I.S. Gradshteyn and I. M. Ryzhik, {\it Table of Integrals, Series and Products},  (Academic Press is an imprint of Elsevier, USA, 2007).
\bibitem{dmft1} Y. Tada, R. Peters, M. Oshikawa, A. Koga, N. Kawakami, and S. Fujimoto, Phys. Rev. B {\bf 85}, 165138 (2012).
\bibitem{dmft2} H.-S. Nguyen, M.-T. Tran, Phys. Rev. B {\bf 88}, 165132 (2013).
\bibitem{dmft3}  T. Yoshida and N. Kawakami, Phys. Rev. B {\bf 94}, 085149 (2016).
\end{thebibliography}
\end{document}